\def\MPL #1 #2 #3 {Mod.~Phys.~Lett.~{\bf#1},\  #2 (#3)}
\def\NPB #1 #2 #3 {Nucl.~Phys.~{\bf#1},\  #2 (#3)}
\def\PLB #1 #2 #3 {Phys.~Lett.~{\bf#1},\  #2 (#3)}
\def\PR #1 #2 #3 {Phys.~Rep.~{\bf#1},\ #2 (#3)}
\def\PRD #1 #2 #3 {Phys.~Rev.~{\bf#1},\  #2 (#3)}
\def\PRL #1 #2 #3 {Phys.~Rev.~Lett.~{\bf#1},\  #2 (#3)}
\def\RMP #1 #2 #3 {Rev.~Mod.~Phys.~{\bf#1},\  #2 (#3)}
\def\ZP #1 #2 #3 {Z.~Phys.~{\bf#1},\  #2 (#3)}
\def\IJMP #1 #2 #3 {Int.~J.~Mod.~Phys.~{\bf#1},\  #2 (#3)}

\newcommand{\be}{\begin{equation}}
\newcommand{\ben}{\begin{subequations}}
\newcommand{\een}{\end{subequations}}
\newcommand{\beq}{\begin{eqalignno}}
\newcommand{\eeq}{\end{eqalignno}}
\newcommand{\ee}{\end{equation}}

\def\ibid{{\it ibid.}}

\def\em{e^-}

\def\ptmin{p_T^{\gam~{\rm min}}}
\def\thetamin{\theta_{\rm min}}
\def\thetae{\theta_e}
\def\thetagam{\theta_\gam}
\def\emax{E^{\rm max}}

\def\wtil{\widetilde}

\def\dmchi{\Delta m_{\tilde \chi_1}}
\def\mchi{m_{\tilde\chi_1}}
\def\mzstar{m_{Z^\star}}
\def\sign{{\rm sign}}

\def\rts{\sqrt s}

\def\h{h}

\def\lam{\lambda}

\def\eg{{\it e.g.}}
\def\etal{{\it et al.}}
\def\epem{e^+e^-}
\def\mupmum{\mu^+\mu^-}

\def\ee{e^+e^-}

\def\lsim{\mathrel{\raise.3ex\hbox{$<$\kern-.75em\lower1ex\hbox{$\sim$}}}}
\def\gsim{\mathrel{\raise.3ex\hbox{$>$\kern-.75em\lower1ex\hbox{$\sim$}}}}
\def\@versim#1#2{\vcenter{\offinterlineskip
        \ialign{$\m@th#1\hfil##\hfil$\crcr#2\crcr\sim\crcr } }}
\def\zstar{Z^\star}

\def\slash#1{#1\hskip-6pt/\hskip2pt}

\def\etmiss{\slash E_T}

\def\ie{{\it i.e.}}

\def\gam{\gamma}

\def\nsd{N_{SD}}
\def\anti{\overline}
\def\pbi{~{\rm pb}^{-1}}
\def\fbi{~{\rm fb}^{-1}}
\def\fb{~{\rm fb}}
\def\pb{~{\rm pb}}

\def\mev{\,{\rm MeV}}
\def\gev{\,{\rm GeV}}
\def\tev{\,{\rm TeV}}

\def\wt{\widetilde}

\def\rta{\rightarrow}
\def\mhalf{m_{1/2}}
\def\gl{\wt g}
\def\mgl{m_{\gl}}

\def\sq{\wt q}

\def\msq{m_{\sq}}

\def\hl{h^0}
\def\hh{H^0}
\def\ha{A^0}

\def\hpm{H^{\pm}}

\def\mhh{m_{\hh}}
\def\mha{m_{\ha}}

\def\mhpm{m_{\hpm}}
\def\tanb{\tan\beta}

\def\mt{m_t}
\def\mb{m_b}
\def\mz{m_Z}

\def\mgut{M_U}
\def\mstring{M_S}
\def\mth{m_{3/2}}
\def\delgs{\delta_{GS}}

\def\wpm{W^{\pm}}

\def\chitil{\wt\chi}
\def\cnone{\wt\chi^0_1}
\def\cntwo{\wt\chi^0_2}
\def\cnthree{\wt\chi^0_3}
\def\cnfour{\wt\chi^0_4}

\def\mcnone{m_{\cnone}}
\def\mcntwo{m_{\cntwo}}
\def\mcnthree{m_{\cnthree}}
\def\mcnfour{m_{\cnfour}}
\def\h{h}

\def\cpone{\wt \chi^+_1}
\def\cmone{\wt \chi^-_1}
\def\cptwo{\wt \chi^+_2}
\def\cmtwo{\wt \chi^-_2}
\def\cpmone{\wt \chi^{\pm}_1}
\def\cmpone{\wt \chi^{\mp}_1}
\def\cpmtwo{\wt \chi^{\pm}_2}
\def\cmptwo{\wt \chi^{\mp}_2}
\def\mcpone{m_{\cpone}}
\def\mcmone{m_{\cmone}}
\def\mcpmone{m_{\cpmone}}
\def\mcpmtwo{m_{\cpmtwo}}

\def\kpr{K^\prime} 

\documentstyle[12pt,equations]{article}
\def\ie{{\it i.e.}}
\def\etal{{\it et al.}}
\def\9{\phantom 0}      
\renewcommand\linebreak{\unskip\break} 
\begin{document}
\input psfig.sty
\newlength{\captsize} \let\captsize=\small 
\newlength{\captwidth}                     

%
\font\fortssbx=cmssbx10 scaled \magstep2
\hbox to \hsize{
%
%
$\vcenter{
\hbox{\fortssbx University of California - Davis}
}$
\hfill
$\vcenter{
\hbox{\bf UCD-96-20} 
\hbox{\bf MADPH-95-951} 
\hbox{July 1996}
\hbox{Revised: February 1999}
}$
}

%
\medskip
\begin{center}
\bf
A NON-STANDARD STRING/SUSY SCENARIO AND ITS PHENOMENOLOGICAL IMPLICATIONS
\\
\rm
\vskip1pc
{\bf C.-H. Chen$^a$, M. Drees$^{a,b}$, and J.F. Gunion$^a$}\\
\medskip
\small\it
$^a$Davis Institute for High Energy Physics, 
University of California,  Davis, CA 95616\\
$^b$Physics Department, University of Wisconsin, Madison, WI 53706\\
\end{center}

\begin{abstract}
We investigate the phenomenology of an orbifold string model in which
supersymmetry breaking is dominated by the overall `size' modulus
field and all matter fields are in the untwisted sector.
The possibly close degeneracy of the lightest neutralino and chargino and
the possibly small splitting between the gluino and chargino/LSP
mass imply that discovery of supersymmetry at future colliders
could be more challenging than anticipated. Specialized search strategies
and particular detector features could play an important role.
For preferred model parameter choices, the phenomenology of dark
matter in the universe is significantly modified.
\end{abstract}

\section{Introduction}

\indent\indent 
In the conventional approach to supersymmetry breaking,
a more or less standard set of boundary conditions at the GUT
scale, $\mgut$, has been most thoroughly studied. These are:
a universal gaugino mass, $\mhalf$; a common scalar mass $m_0$;
and a universal value for the soft $A$ parameters.
The effects of non-degeneracy among the scalar masses
have recently been explored in many papers
\cite{nonu}.  No systematic
study of models with non-universal gaugino masses has appeared.
In this paper, we explore one such model, motivated by the string
theory picture, in which supersymmetry breaking is dominated by the
overall `size' modulus field, as described in Ref.~\cite{ibanez}.
Models with multiple modulus fields are
considered in Ref.~\cite{ibanezii}. 
In this latter paper, it is found that if tachyonic masses
that could cause charge
and color breaking are not allowed, then the soft-SUSY-breaking
boundary conditions cannot be such as to greatly
distort results based on assuming
that all the moduli fields participate in SUSY breaking
equally, as is implicit when only the overall modulus
field is employed. 
The boundary conditions that result when SUSY breaking
is dominated by the overall size modulus field
lead to an interesting and unusual
SUSY phenomenology that differs substantially from that
appropriate to universal boundary conditions in such diverse
areas as cold dark matter in the universe and direct SUSY
search strategies/difficulties.

In the context of string model scenarios in which
all the moduli fields participate equally through an overall modulus field,
non-universality among the gaugino masses at $\mgut$ is not typical.
This is because the only tree-level contributions to the gaugino masses
are those originating from the dilaton field, and these
are automatically universal. 
Non-universal contributions to the gaugino masses arise only at
one-loop. Thus, significant non-universality is only possible
when SUSY breaking is not dominated by the dilaton, but rather
by the overall modulus field.  An additional motivation for considering 
a model
with modulus (as opposed to dilaton) dominated SUSY breaking is the fact
that essentially the whole of dilaton-dominated parameter space is
excluded by the requirement that the standard SUSY vacuum
should be deeper than the charge and color breaking minima \cite{clm}.

In the notation of Ref.~\cite{ibanez}, SUSY breaking becomes moduli-dominated 
in the $\sin\theta\rta 0$ limit, where $\theta$ is the goldstino
angle and $\tan\theta$ measures the relative amount of SUSY
breaking due to the dilaton field, $S$, relative to the overall size modulus
field, $T$.
In the $\sin\theta\to 0$ limit, the gaugino masses at $\mgut$
arise entirely from one-loop threshold and Kahler potential 
corrections. The masses are both small and typically very non-universal.

Although it is possible to take $\sin\theta\to0$ 
in a general Calabi-Yau model, there is only one simple orbifold model in which
$\sin\theta$ can be taken to be sufficiently near zero to avoid dilaton
dominance and the resulting approximate universality of gaugino masses.
This is the model (called the O-II scenario) in which
all matter fields have modular weight $n=-1$, \ie\ lie in the untwisted sector.
(If any $n\neq -1$, then $\sin^2\theta\geq 1/2$ is required
in order to avoid a negative mass squared for some squark
or slepton.) The $\sin\theta\rta 0$ limit of the O-II model 
is analogous to the effective large $T$ limit of a Calabi-Yau model.
We focus on the $\sin\theta\to0$ limit of the O-II orbifold model
since it is only for orbifold models that the required
one-loop Kahler potential and threshold corrections have been computed.
Although, this model is only one of many theories that yield
non-universal gaugino masses, its phenomenology will
provide a number of very valuable lessons and comparisons
to the phenomenology typical of models with universal boundary conditions.

In the O-II orbifold model,
one obtains the following boundary conditions at the string scale, $\mstring$,
in the limit of $\sin\theta\to0$:
\begin{equation}
\begin{array}{l}
M_3^0=1.0\sqrt 3 \mth[-(3+\delgs)K\eta] \\
M_2^0=1.06\sqrt 3\mth[-(-1+\delgs)K\eta] \\
M_1^0=1.18\sqrt 3\mth[-(-33/5+\delgs)K\eta] \\
m_0^2=\mth^2[-\delgs\kpr] \\
A_0=0
\end{array}
\label{bcs}
\end{equation}
where we shall employ the one-loop numerical estimates of Ref.~\cite{ibanez},
$K=4.6\times 10^{-4}$ and $\kpr\equiv 1/(24\pi^2Y)=10^{-3}$.
(The 0 subscript or superscript indicates the $\mgut$-scale value.)
In the above, $\delgs$ is the Green-Schwarz mixing parameter,
which preferably lies in the range 0 to $-5$. In the specific
O-II model considered, $\delgs$ would be a negative integer, with
$\delgs=-4,-5$ preferred.
In Eq.~(\ref{bcs}), $\eta=\pm1$, corresponding to $\sin\theta\rta0$
with $\cos\theta=\pm1$ (\ie\ $\theta\to 0,\pi$). 

In Ref.~\cite{ibanez}, two sources for the $B$
parameter were considered, labelled by $B_Z$ and $B_\mu$.
Here, $B\mu$ is the
coefficient of the $H_1H_2$ mixing term in the scalar Higgs
sector potential. (We employ the conventions for $B$ and $\mu$
of Ref.~\cite{hhg}.) The source of $B_Z$ is the Giudice-Masiero
mechanism \cite{gmmech}.
It was stated that only $B_\mu$ could be present
in orbifold models.  More recently \cite{ibanezii}, it was
realized that $B_Z$ could also be present, and, as well,
a third type of $B$ contribution, $B_\lam$, was discussed.
In general, there could be a mixture of all three.  Here,
we focus on just $B_Z$ and $B_\mu$.
A somewhat uncertain (in the sense
that many additional approximations are made beyond
those required for Eq.~(\ref{bcs})
prediction for $B_\mu$ in the O-II model is:
\begin{equation}
B^0_\mu=\mth(-1-(1-\delgs\kpr)^{-1/2}\eta)\,,
\label{bform}
\end{equation}
If one were to adopt this prediction for $B$, then the value of $B$
would be extremely large for $\eta=+1$, and only $\eta=-1$
can possibly be phenomenologically relevant. For this choice,
\begin{equation}
B^0_\mu\simeq -{1\over 2}m_0\sqrt{-\delgs\kpr}
\label{bapprox}
\end{equation}
and $B^0$ would typically be much smaller than $m_0$ for the value of $\kpr$
we employ. The result for $B^0_Z$ is very different:
\begin{equation}
|B^0_Z||\mu_0|=(m_0^2+|\mu_0|^2)\,,
\label{bgmmech}
\end{equation}
which corresponds to $\tanb=1$ at $\mgut$.
Because of the large number of possibilities for $B$ 
we shall leave $B^0$ as a free parameter; it turns out to be closely
related to $\tanb$, 
where $\tanb=v_2/v_1$ is the ratio of the vacuum expectation values
of the neutral components of the Higgs doublet fields responsible for giving
mass to the up- and down-types quarks, respectively. 
We find that $\tanb$ (at $\mz$) near 1 is required
for  pure $B_Z$ and that very large $\tanb$ is needed for consistency
with pure $B_\mu$.

If the model prediction for the $B$ parameter is ignored,
the sign of $\eta$ becomes physically irrelevant since the 
overall sign of the GUT scale gaugino 
masses in Eq.~(\ref{bcs}) can be rotated away
(since $A_0=0$) by an appropriate overall
phase choice for the gaugino fields. (However, the opposite
sign of $M_3^0$ relative to $M_{1,2}^0$ for $|\delgs|<3$ is physically
relevant; it impacts the running of the $A$ parameters and of $B$.)

We shall consider these boundary conditions within the context
of the minimal supersymmetric model (MSSM) with exactly two Higgs doublets.
This context is motivated by the fact that it is
only for exactly two doublets (plus possible singlets) that
the coupling constants unify without intermediate scale matter.
However, it must be noted that the scale $\mgut$ at which
the coupling constants unify is substantially
below the string scale, $\mstring$, 
at which the above boundary conditions naively apply.
Thus, we shall be implicitly assuming that there is
some effect, such as chiral fields in the spectrum between $\mgut$
and $\mstring$, that compensates for this discrepancy.

Taking $B^0$ to be a free parameter in addition to $m_0$ and $\delgs$, 
we evolve down to scales below a TeV and fix the superpotential parameter $\mu$ 
(which appears in the $\mu\hat H_1\hat H_2$ superpotential term) by requiring
that electroweak symmetry breaking (EWSB) gives the correct value of $\mz$.
The sign of $\mu$ remains undetermined.
In practice, it is more convenient to trade the parameter $B^0$
for the parameter $\tanb$.
The top and bottom quark Yukawa couplings are constrained to yield the observed
values of $\mt$ and $\mb$, which we take to be $\mt(\mt)=165\gev$
and $\mb(\mb)=4.25\gev$. We do not insist on $b-\tau$ Yukawa unification.
The free parameters of the model are thus:
\begin{equation}
m_0\,,\quad
\delgs\,,\quad
\tanb\,,\quad \sign(\mu)\,.
\end{equation}
We will often consider a fixed value for $m_0$ and plot results
as a function of $\tanb$ and/or $\delgs$. For fixed choices of $m_0$, $\delgs$
and $\sign(\mu)$, $B^0$ can be viewed as a function of $\tanb$
and its value can be compared to the rough model predictions
of Eqs.~(\ref{bapprox}) and (\ref{bgmmech}). 
We shall return to this comparison shortly.

It is useful to summarize the behavior and magnitude
of the $M_i^0$ as a function of $-\delgs$.
From Eq.~(\ref{bcs}) we find:
\begin{itemize}
\item
When $\delgs\to0$ at fixed $m_0$, the $|M_i|$ grow roughly as 
\begin{equation}
\left[|M_1^0|,|M_2^0|,|M_3^0|\right]\sim\left[{33\over 5},1,3\right]
\times\left({\sqrt 3 K m_0\over \sqrt{-\delgs \kpr}}\sim {0.025 m_0\over
\sqrt{-\delgs}}\right)\,.
\label{smalld}
\end{equation}
\item 
For $\delgs=-4,-5$, as possibly
preferred in the O-II model, $m_0\gg |M_1^0|\gg|M_2^0|\gg|M_3^0|$.
For $\delgs=-3$, $|M_3^0|=0$.
\item
If $-\delgs$ is large (roughly $-\delgs\gsim 30-40$), then the
$|M_i^0|$ become approximately universal, with 
\begin{equation}
|M_i^0|\sim \sqrt{-\delgs}{\sqrt 3 Km_0\over \sqrt{\kpr}}
\sim0.025 m_0 \sqrt{-\delgs}\,.
\label{larged}
\end{equation}
\end{itemize}
The first important point to note 
is that, unless $\delgs$ is extremely large ($-\delgs\gsim 100$)
or very small ($-\delgs\lsim 0.001$),
the (tree-level) value of $m_0$ is very much larger than any of the 
$|M_i^0|$; this is basically due to the fact that
$K$ and $\kpr$ are similar in size and have typically small
one-loop magnitudes. Consequently,
the squarks, sleptons and heavier Higgs bosons ($\hh$,
$\ha$ and $\hpm$, with $\mhh\sim\mha\sim\mhpm$) in this
model will be much heavier than the gauginos. Further, the $M_i^0$ themselves
are very non-universal unless $|\delgs|$ is large.
In particular, for any moderate choice of $\delgs$, $|M_1^0|\gg |M_2^0|$,
implying that the lightest chargino and lightest neutralino
will both be wino-like and nearly degenerate in mass.
This will have important phenomenological implications.
Values of $\delgs\sim -3$, \ie\ near the zero of $|M_3^0|$,
will be physically disallowed by the requirement that the gluino
cannot be the lightest supersymmetric particle.

\begin{figure}[htb]
\begin{center}
\centerline{\psfig{file=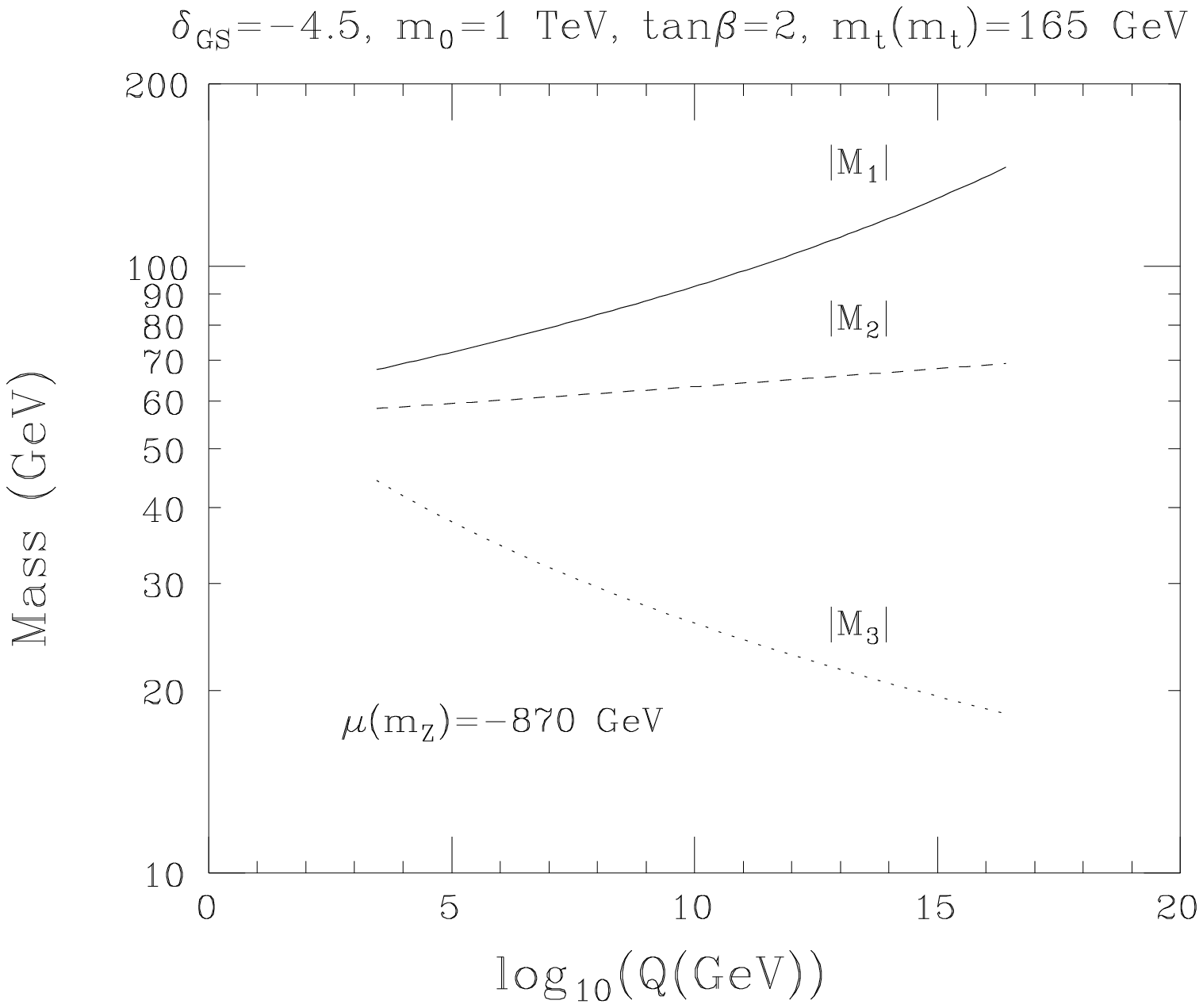,width=12cm}}
\bigskip
\begin{minipage}{12.5cm}       
\caption{Running values for the $|M_i|$ as a function of the scale $Q$,
taking $m_0=1\tev$, $\tanb=2$, $\delgs=-4.5$ $\mt(\mt)=165\gev$,
and $\sign(\mu)=-1$.
}
\label{runningmasses}
\end{minipage}
\end{center}
\end{figure}

\begin{figure}[htb]
\begin{center}
\centerline{\psfig{file=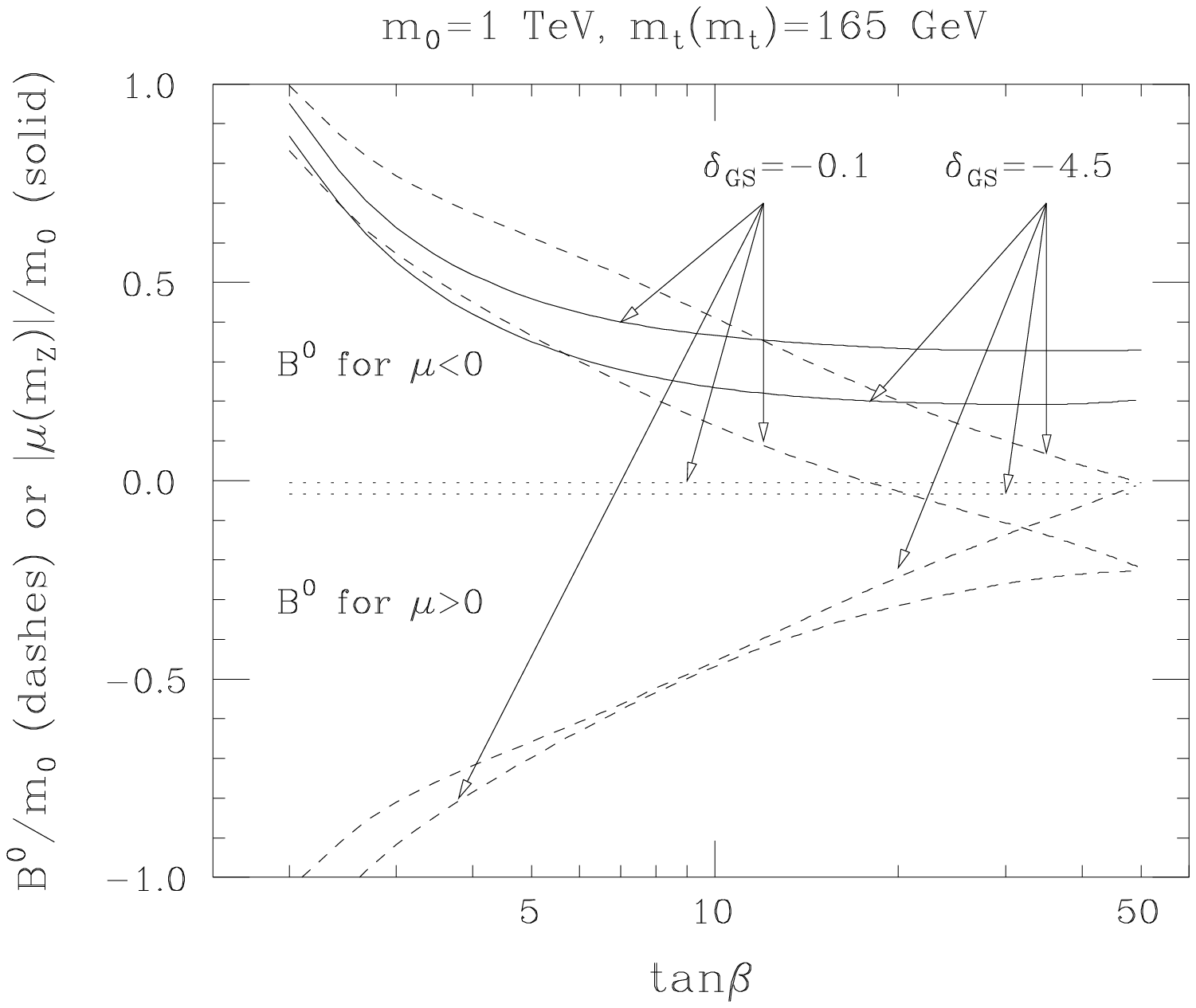,width=12cm}}
\bigskip
\begin{minipage}{12.5cm}       
\caption{Values of $B^0/m_0$ (dashes) and $|\mu(\mz)|/m_0$ (solid)
as a function of $\tanb$ for $\delgs=-0.1$ and $-4.5$,
taking $m_0=1\tev$ and $\mt(\mt)=165\gev$. Results for $|\mu(\mz)|$ 
are the same for $\mu>0$ and $\mu<0$; $B^0$ depends upon the sign of $\mu$,
with $B^0>0$ being favored for $\mu<0$ and vice versa. The two horizontal
dotted lines are the values of $B^0$ 
predicted by Eq.~(\protect\ref{bapprox}).
}
\label{par}
\end{minipage}
\end{center}
\end{figure}

For $\delgs=-4,-5$ or thereabouts, $|M_1|\gg|M_2|\gg|M_3|$ at $\mgut$
and the $|M_i|$ approach one another as one evolves down to $\mz$.
This is illustrated in Fig.~\ref{runningmasses}.  Fig.~\ref{par} illustrates
the $B^0=B(\mgut)$ and $\mu(\mz)$ parameter values
as a function of $\tanb$ for $\delgs=-0.1$ and $-4.5$. (Results
for still larger $|\delgs|$ are rather close to those plotted
for $\delgs=-4.5$.) Both $B$ and $\mu$ evolve rather slowly as
a function of scale. Note that $|\mu(\mz)|$ is independent
of the sign of $\mu$, but that $|B^0|$ is not.  The sign of $B^0$
is generally opposite that of $\mu$ for correct electroweak symmetry breaking.
Except for $\mu>0$ and small $|\delgs|$,
the $B^0$ required by EWSB (dashed lines) crosses the 
approximate model prediction of Eq.~(\ref{bapprox}) (indicated by the dotted
lines) at high $\tanb$ before $\tanb$ exceeds the $\tanb\lsim 50$ limit imposed
by perturbativity for the Yukawa couplings.  Clearly, the $B_\mu$ prediction
of the O-II model for $B^0$ is generally consistent 
with the requirements of EWSB only if $\tanb$ is large. 
From the plots of $B^0$ and $|\mu|$ in Fig.~\ref{par}, it is also apparent
that pure $B_Z$ is only possible if $\tanb$ is near 1.

We note that the existence of solutions with $\mu^2>0$ (as required for
correct electroweak symmetry breaking) at large $\tanb$ is rather
sensitive to the value of $\mt(\mt)$.  This is because the value of $\mu^2$
required for EWSB drops rapidly (see Fig.~\ref{par}) as $\tanb$ increases.
For values of $\mt(\mt)\lsim 160\gev$
(\ie\ not much lower than the $165\gev$ value chosen here),
$\mu^2<0$ is required for correct EWSB at scale $\mz$ if $\tanb$ is large.
The results presented in this paper employ one-loop renormalization
group equations; the full two-loop equations for the entire system
of RGE equations are very difficult to implement. 
It is conceivable
that $\mu^2$ remaining positive out to large $\tanb$ for $\mt(\mt)=165\gev$
could be altered in the full two-loop implementation.

\begin{figure}[htb]
\begin{center}
\centerline{\psfig{file=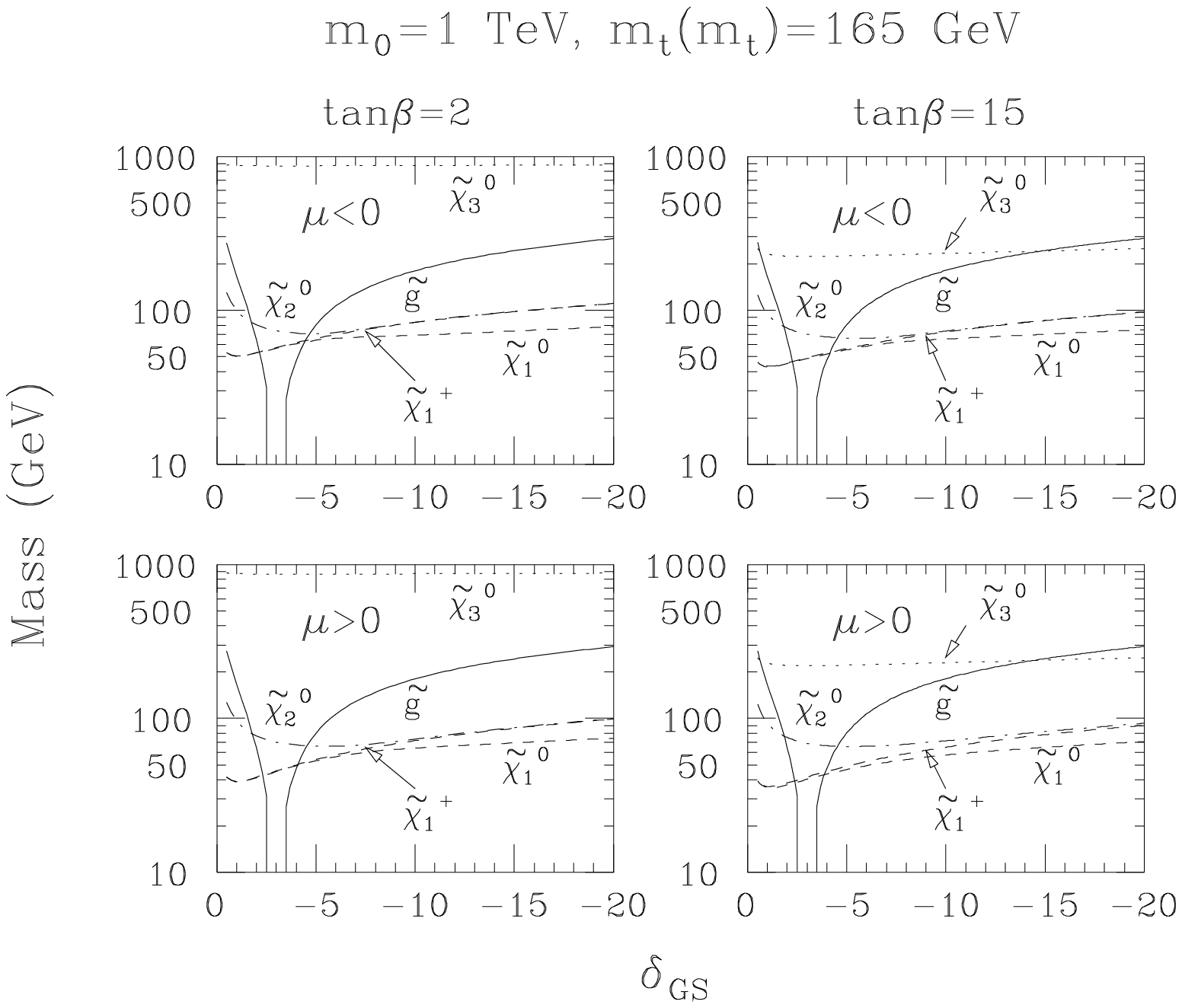,width=12cm}}
\bigskip
\begin{minipage}{12.5cm}       
\caption{Masses for the $\cnone$, $\cntwo$, $\cnthree$, $\cpone$,
and $\gl$ are plotted as a function of $\delgs$ at $\tanb=2$ and 15
and for $\mu>0$ and $\mu<0$.
}
\label{inos}
\end{minipage}
\end{center}
\end{figure}

\begin{figure}[htb]
\begin{center}
\centerline{\psfig{file=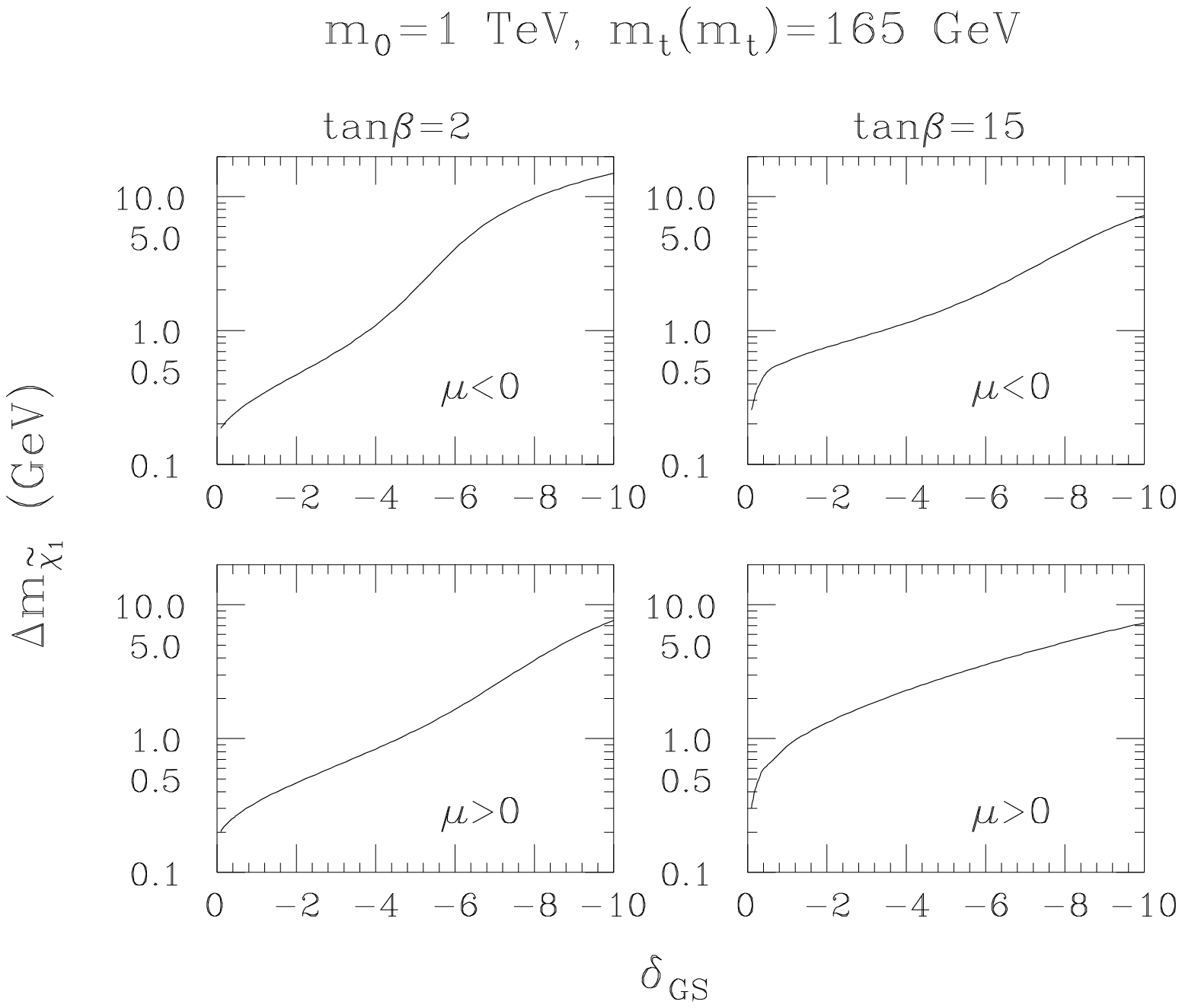,width=12cm}}
\bigskip
\begin{minipage}{12.5cm}       
\caption{The mass splitting $\dmchi\equiv\mcpmone-\mcnone$
as a function of $\delgs$ at $\tanb=2$ and 15 and for $\mu>0$ and $\mu<0$.
}
\label{inosdmchi}
\end{minipage}
\end{center}
\end{figure}

Typical results for the gaugino masses as a function of $\delgs$
are illustrated in Fig.~\ref{inos} for $\mu>0$ and $\mu<0$, taking $m_0=1\tev$
and $\tanb=2,15$. Note that a large value of $m_0$ is required 
for $\mcpmone>\mz/2$.  As discussed later, this lower bound 
from LEP-I data continues to apply in the present model. 
We observe that for $|\delgs|<5$
the $\cnone$ and $\cpone$ are extremely degenerate.
When this near degeneracy is present we will use the notation:
\begin{equation}
\dmchi\equiv \mcpmone-\mcnone,\quad \mchi\equiv \mcpmone\simeq\mcnone\,.
\label{degdef}
\end{equation}
The degeneracy slowly eases as $\tanb$ increases. 
Fig.~\ref{inosdmchi} displays the $\cpmone$---$\cnone$
mass splitting in more detail. In obtaining precise values
for $\dmchi$ it is important to include loop corrections,
the only significant such corrections being from gauge-Higgs loops.
We employed the results of Refs.~\cite{mngy,pappierce}.
Fig.~\ref{inosdmchi} shows that
$\dmchi$ can be as small as $150\mev$
at small $|\delgs|$, if $\tanb=2$. For $\delgs\sim -5$, $\dmchi$
is $\lsim 1-2\gev$ in all cases. Values of $\dmchi\gsim 10\gev$
are only achieved for $|\delgs|>10-15$, depending upon $\tanb$
and $\sign(\mu)$. From Fig.~\ref{inos},
we also observe that for values of $\delgs$ in the vicinity of $-3$,
the gluino becomes the lightest supersymmetric particle.\footnote{We
always discuss and plot $\mgl(pole)$. Thus, for example,
even though $|M_3|$ lies below $|M_2|$ in Fig.~\protect\ref{runningmasses},
the pole value of $\mgl$ is greater than $\mcnone$.
For the parameters of the plot, $\mgl(pole)\sim 66\gev$ while
$\mcnone\sim 62.4\gev$.}
Such values of $\delgs$ are excluded by
cosmological arguments which imply that the LSP cannot be colored.
Note that the other gauginos also have their minimum masses
in the vicinity of this disallowed region. As noted
in Eqs.~(\ref{smalld}) and (\ref{larged}), as $-\delgs\to 0$
or for large $-\delgs$ the $|M_i^0|$ increase away from their minimum values.
Finally, we note that for very small $|\delgs|$ the ratio $\mgl/\mcpmone$
reaches values as high as 6 to 7, substantially above the value 
$\sim 3$ typical of a model with universal $|M_i^0|$ at $\mgut$.

\begin{figure}[htb]
\begin{center}
\centerline{\psfig{file=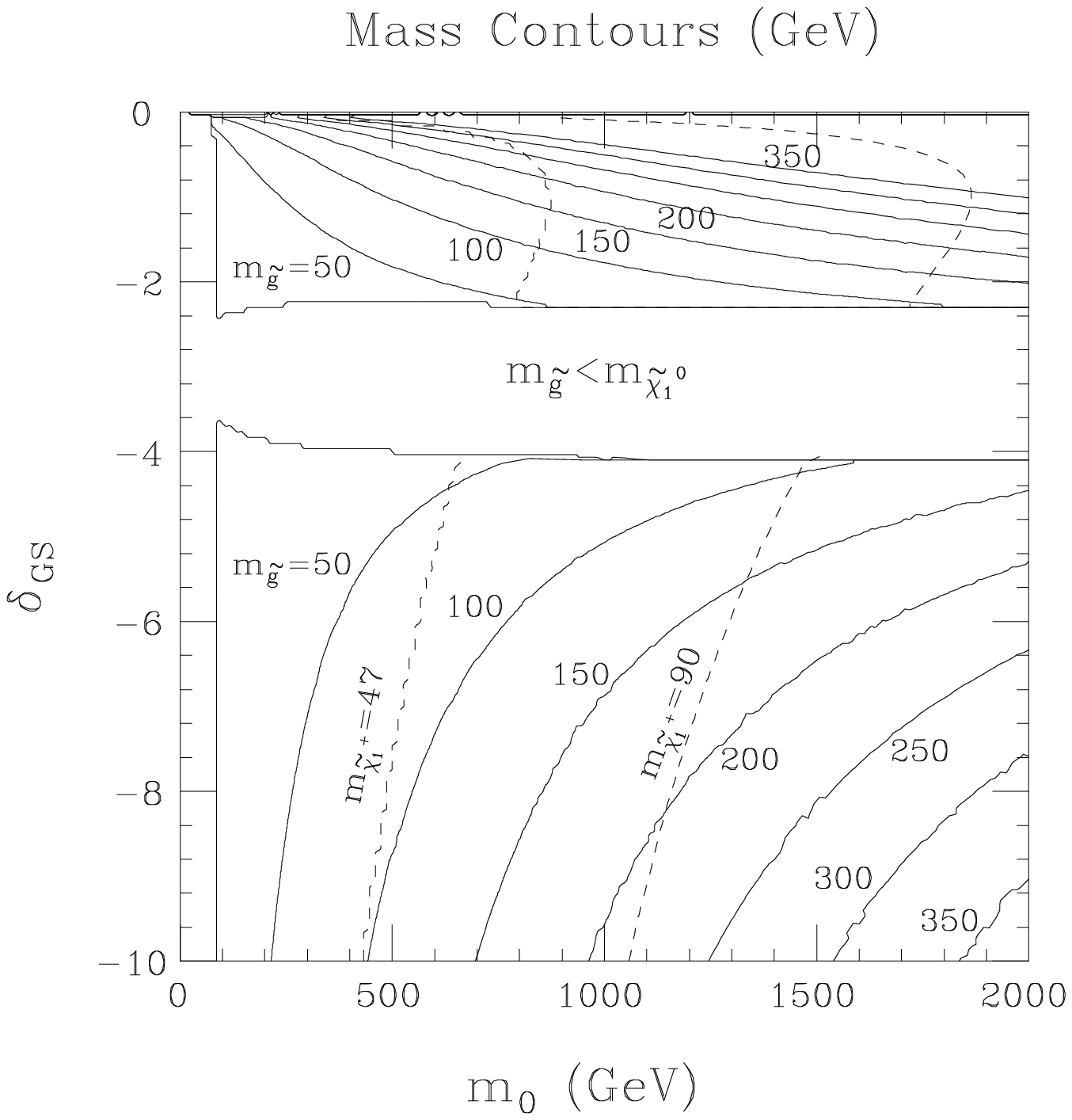,width=11cm}}
\bigskip
\begin{minipage}{12.5cm}       
\caption{Gluino mass contours in $\delgs$--$m_0$ parameter space.
The gluino contours are independent of $\tanb$ and $\sign(\mu)$.
Also shown are the corresponding $\mcpone=47\gev$ and $90\gev$ contours
for $\tanb=2$ and $\mu<0$.
The $\cpone$ contour depends only weakly on $\tanb$ and $\sign(\mu)$.
}
\label{mglcontours}
\end{minipage}
\end{center}
\end{figure}

The near equality of $\mcpone$ and $\mcnone$ at lower $|\delgs|$
follows from the fact that $|M^0_1|$ is large compared to $|M^0_2|$ at $\mgut$,
which implies that even though $|M_1|$ falls towards $|M_2|$
as the scale decreases, $|M_1|>|M_2|$ at $\mz$. As a result,
for such $\delgs$, the $\cnone$ and the $\cpone$
are both primarily winos and thus have very similar mass.

In contrast to the $\cpone$, the $\cntwo$ is never especially
degenerate with the $\cnone$. At small $|\delgs|$, $\mcntwo\sim |M_1|$
is substantially above $\mcpone\sim\mcnone\sim |M_2|$ at scale $\mz$.
At high $|\delgs|$, as the $|M_i^0|$ approach universality, one
approaches the more familiar situation where the $\cntwo$ and $\cpone$
are both winos and $\mcntwo\sim\mcpone>\mcnone$. Also shown
in Fig.~\ref{inos} is $\mcnthree$.  Because the $\cpmtwo$, $\cnthree$
and $\cnfour$ are all primarily higgsino in nature, they will have
similar mass ($\sim |\mu|$).

Additional perspective on masses is provided by Fig.~\ref{mglcontours}
where we give contours for $\mgl=50$ to $350$ in steps of $50\gev$
in $\delgs$--$m_0$ parameter space. The $\mgl$ contours
are independent of $\tanb$ and $\sign(\mu)$.
Also shown are the contours
for $\mcpone=47\gev$ and $90\gev$, for $\tanb=2$ and $\mu<0$.
Parameter space points to the left
of the $47\gev$ contour are excluded by LEP-I data, implying
that $\mgl$ must lie above about $50\gev$. The gap region is that
excluded by requiring that the $\gl$ not be the LSP. 
Note that $\mgl/m_0$ is small when $|\delgs|$ is not large.
Thus, for example, if we assume that naturalness demands
that $m_0$ lie below about $2\tev$, the maximum $\mgl$ that
can be achieved along the $\mgl\sim\mchi$ [see Eq.~(\ref{degdef})]
border is of order $150\gev$. Even in the extreme $\delgs\sim -10$,
$m_0=2\tev$ corner of the plot, $\mgl\sim 375\gev$. 
Large $\mgl$ values can only be achieved by taking either very
small or very large $-\delgs$,
keeping $m_0$ fixed. Despite the generally small size of $\mgl$, 
we will see that $\gl$ detection at a hadron collider is challenging
along the $\mgl\sim\mchi$ boundary.
As one moves away from the $\mgl\sim \mchi$ boundary,
discovery of the $\gl$ becomes easier, eventually approaching
expectations for the canonical universal boundary condition scenarios.

The masses for all sleptons, heavy Higgs bosons,
and squarks are of the order of $m_0$.  As seen in Figs.~\ref{inos} and 
\ref{mglcontours}, unless $|\delgs|$ is very large or very small,
$m_0$ is much larger than the gaugino masses.  Consequently,
for the $\sin\theta\to0$ O-II model being considered it is
likely that the gauginos
will be the most abundantly produced SUSY particles,
and they would probably provide
the first observed SUSY signals. The next section is devoted
to a discussion of the search strategies required, emphasizing the
difficulties that arise in $\cpmone$ detection at an $\epem$ collider
when the $\cpone$ and $\cnone$ are closely
degenerate and in discovering the gluino at
a hadron collider when $\mgl$ is close to $\mchi\equiv\mcpmone\simeq\mcnone$.

\section{Phenomenology}

In this section we discuss SUSY
discovery strategies for the $\sin\theta\to0$ O-II model. Because
of the special features of the mass spectrum, SUSY discovery
can be substantially more difficult than in models where the gaugino
masses are universal at $\mgut$. 
Universality at $\mgut$ implies the hierarchy
\begin{equation}
|M_1|\sim {1\over 2} |M_2|\sim {1\over 6} |M_3|< |\mu|\,,
\end{equation}
at scale $\mz$.
In this case, the $\cnone$ LSP is a bino and its mass, $\sim |M_1|$,
is significantly lower than the masses of
the $\cpone$ and $\cntwo$, $\sim |M_2|$, which
is substantially below $\mgl\sim |M_3|$.
As a consequence, the $\gl$ typically decays to the $\cpone$
or $\cnone$ plus a pair of energetic quark jets
and the $\cpone$ decays to the $\cnone$ by emitting a $\ell\nu$
or $q^\prime \anti q$ pair with significant energy.  The substantial
energy carried by the decay products implies that production
of the lightest chargino and of the gluino
will be associated with both energetic jets/leptons and substantial missing
energy, a combination that is generally easily separated from
backgrounds at an $\epem$ or hadron collider operating at high enough energy
and luminosity. 

In the O-II model, SUSY detection need not be so
straightforward. We have already noted that it is natural for
$\dmchi$ to be sufficiently small that the final particle(s) (quarks,
pion(s) or lepton) emitted when the
$\cpmone$ decays to $\cnone$ are very soft and not easily
observed. We will also find that the $\cpmone$'s lifetime
is not likely to be sufficiently long that it will appear
as a stable particle track in the detector --- short tracks in a 
vertex detector are, however, a distinct possibility.
Less automatic, but also possible, is degeneracy of $\mgl$
with $\mchi\equiv\mcpmone\simeq\mcnone$. The jets from $\gl$ decay
to $\cpmone$ or $\cnone$ would then be very soft and difficult
to detect.

At an $\epem$ collider, small $\dmchi$ leads to difficulty in detecting
$\epem\to \cpone\cmone$. Indeed, the techniques discussed for
isolating the $\cpone\cmone$ signal at LEP-II (\ie\
above the $Z$ pole) have good efficiency only for $\dmchi\gsim
10\gev$.\footnote{In a recent paper \cite{l3recent}, the L3
collaboration mentions a specialized technique employed
at $\rts\sim 130-140\gev$ for retaining some $5-10\%$ efficiency down
to $\dmchi=5\gev$, but details are not given.}
Most likely, it would be necessary
to employ other reactions to first discover SUSY. One possibility
is the much smaller $\epem\to\cnone\cntwo,\cntwo\cntwo$
production processes, in which the $\cntwo$ would generally
yield energetic and visible decays products, given that
the smallest $\mcntwo-\mcnone$ mass difference values are of order
$5-10\gev$ (for $\delgs$ between $-6$ and $-9$). Another is
$\epem\to\gam\cpone\cmone$ production in which the nearly or completely
invisible $\cpone\cmone$ pair is tagged by detecting the hard $\gam$.
This latter was investigated in Ref.~\cite{cdgi}, and will be
reviewed shortly.

At a hadron collider, if $\mgl$ is close to $\mchi$,
the softness of the jets in $\gl$ decay implies that 
the usual procedures
for isolating gluino pair production at a hadron collider
by tagging missing energy {\it and} jets may yield a 
rather weak signal. Further, if 
the $\ell$ from $\cpmone$ decay is soft due to small $\dmchi$ then:
(a) the like-sign dilepton signature for $\gl\gl$ production that
derives from $\gl\gl\to \cpmone\cpmone +jets$ 
followed by $\cpmone\cpmone\to \ell^{\pm}\ell^{\pm}\etmiss$ will
be difficult to extract; and, (b) the tri-lepton signature 
for $\cpmone\cntwo$ production will be hard to observe.

Thus, it is clear that the O-II model
leads to a situation where the techniques and prospects for detecting
SUSY must be re-evaluated.

\subsection{Lifetime and Branching Ratios of the {\boldmath$\cpone$}}

\begin{figure}[htb]
\begin{center}
\centerline{\psfig{file=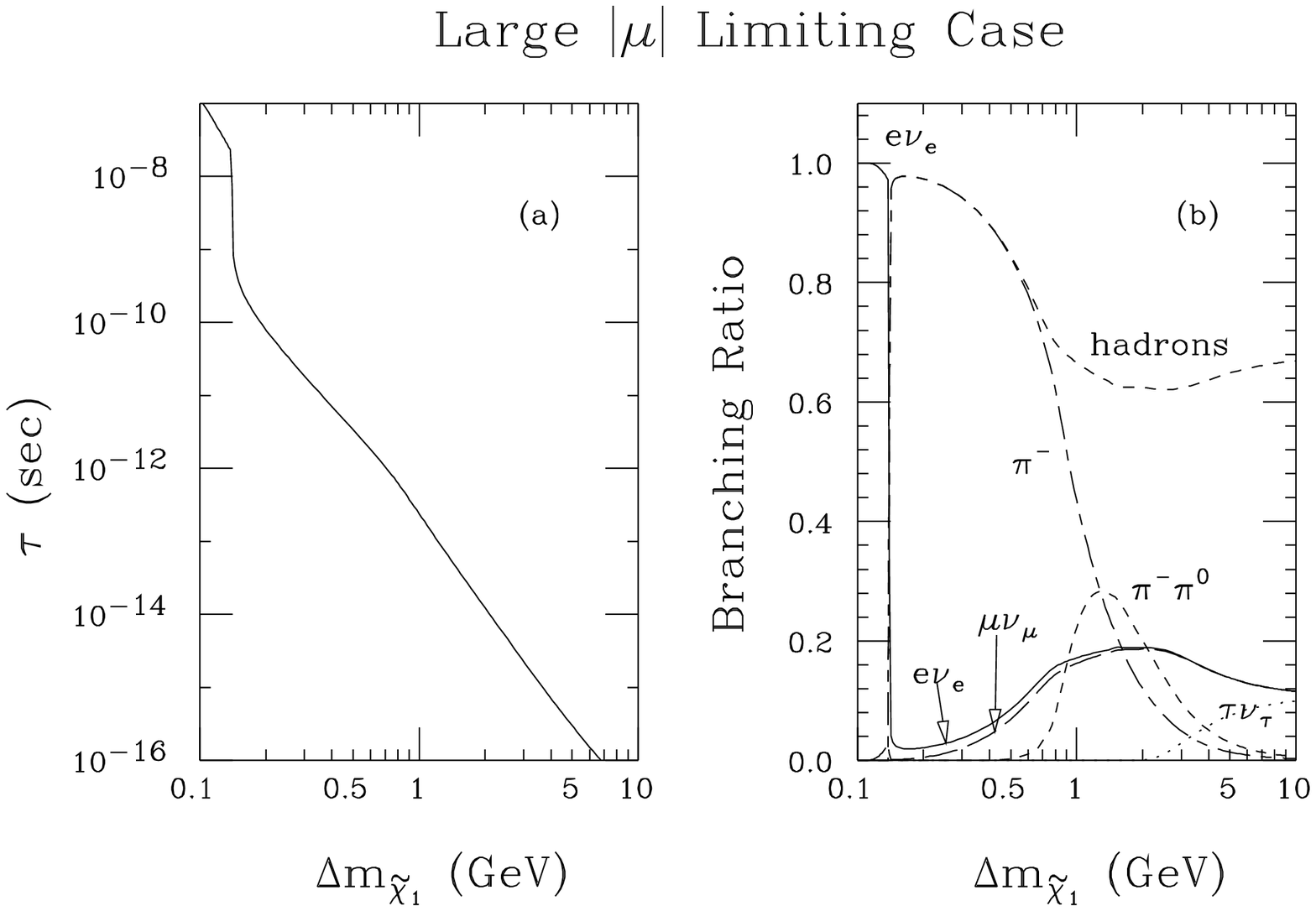,width=12cm}}
\bigskip
\begin{minipage}{12.5cm}       
\caption{We plot the lifetime (a) and the relevant branching ratios (b)
for the $\cmone$ as a function of $\dmchi\equiv \mcmone-\mcnone$.
For $\dmchi< 1.5\gev$,
we explicitly compute and sum the $\pi^-\cnone$, $\pi^-\pi^0\cnone$,
$\pi^-\pi^0\pi^0\cnone$ and $\pi^-\pi^+\pi^-\cnone$ modes. These
merge into and are replaced by a computation of
the $q^\prime\overline q\cnone$ width for $\dmchi>1.5\gev$.
}                                   
\label{lifebrs}
\end{minipage}
\end{center}
\end{figure}

From the above discussion, it is clear that important
ingredients in the phenomenology of SUSY detection in the O-II
model context are the branching ratios and lifetime
of the $\cpmone$. These have been computed using PCAC-style techniques
as described in Appendix A.  We find
that both the branching ratios and the lifetime depend almost
entirely upon the mass difference $\dmchi=\mcpmone-\mcnone$.
Dependence upon $\tanb$ and $\mu$ is minimal.
Results for the lifetime $\tau$ and for the important branching
ratios of the $\cpone$ are plotted in Fig.~\ref{lifebrs}.
For very small $\dmchi$, $\cmone\rta\ell^-\nu_{\ell}\cnone$
($\ell=e,\mu$) is the only kinematically allowed decay mode.
As the mass difference increases, $\cmone\rta\pi^-\cnone$ opens
up and remains dominant for $m_\pi<\dmchi\lsim 1\gev$.
Above that, the $\cnone\to\pi^-\pi^0\cnone$ and three-pion channels become
important. The sum of the one-, two- and three-pion channels merges
into $\cmone\rta q^\prime \overline q \cnone$ at $\dmchi\sim 1.5\gev$.  
For still larger mass difference, $\cmone\rta\tau^-\nu_{\tau}\cnone$ 
becomes kinematically allowed. 

Fig.~\ref{lifebrs}(a) shows that a produced $\cpmone$
will travel distances of order
a meter or more (and thus appear as a heavily-ionizing
track in the vertex detector and the main detector)
if $\dmchi<m_\pi$. 
For $m_\pi<\dmchi<1\gev$, $10~\mbox{cm}>c\tau> 100~\mu$m.
For $c\tau$ near 10 cm, the $\cpmone$ would
pass through enough layers of a typical vertex detector
that its heavily ionizing nature would be apparent.
For $c\tau$ in the smaller end of the above range, one would 
have to look for the single charged pion from the
dominant $\cpmone\to\pi^\pm\cnone$ mode. It emerges
at a finite distance of order $c\tau$ from the vertex
and would have momentum
$p_\pi\sim  \sqrt{\Delta m_{\tilde{\chi}_1}^2-m_\pi^2}$ in
the $\cpmone$ rest frame. The expected impact parameter resolution, $b_{\rm
res}$, of a typical vertex detector (we looked at the CDF Run II
vertex detector with the inner L00 layer in detail~\footnote{We thank
H. Frisch for providing details. The NLC vertex detector will
be similar (R. Van Kooten, private communication).}) 
as a function of momentum is such that
$c\tau/b_{\rm res}>3$ for $\dmchi<1\gev$, with quite large
values typical for $\dmchi<0.5\gev$. Such a high-$b$
pion in association with an appropriate trigger would constitute
a fairly distinctive signal. The magnitude of $\dmchi$ as a function of
$\delgs$ was illustrated in Fig.~\ref{inosdmchi}. For
$|\delgs|\lsim 1-4$, depending upon $\sign(\mu)$ and $\tanb$,
$c\tau$ for the $\cmone$ can be large enough for the heavily-ionizing
$\cpmone$ or the non-zero impact parameter of
the decay pion to be visible in the vertex detector.

\subsection{Constraints from LEP and LEP1.5}

We note that the $\cpmone$
cannot be lighter than $\mz/2$. We have explicitly checked that
the $Z\rta\cpone\cmone$ decays would have been noticed
either as an invisible width contribution or through an enhancement
in the total $Z$ width.
This statement applies for masses up to within a fraction of a GeV of $\mz/2$.
The $Z\rta \cnone\cnone,\cnone\cntwo,\cntwo\cntwo$ 
decays have much smaller widths
(due to the small higgsino component of the $\cnone$ and $\cntwo$)
and do not provide useful direct limits. Implicit limits on $\mcnone$
associated with the $\mcpmone\gsim \mz/2$ limit depend upon $\delgs$.
For small $|\delgs|$ the degeneracy $\mcnone\simeq\mcpmone$
implies that $\mcnone\gsim \mz/2$ as well.  However, for
large $|\delgs|$ the $\cnone$ is significantly lighter than the $\cpmone$,
and $\mcnone< \mz/2$ is allowed. 
At large $|\delgs|$, such that $\dmchi\gsim 5-10\gev$, the LEP1.5
limit of $\mcpmone\gsim 65\gev$ applies (see next section for further
discussion).  The lowest $\cnone$ mass consistent with this limit
for $|\delgs|\leq 20$ is $\mcnone=41\gev$ (corresponding to $m_0=520\gev$
at $\delgs=-20$).

\subsection{SUSY Discovery at $\epem$ Colliders}

Let us begin by considering the neutralino and chargino
pair production cross sections. As the $\rts$ of the machine increases,
the first channels to open up will be those for the lighter eigenstates:
\begin{equation}
\epem\to \gam^*,Z^*\to \cpone\cmone\,,\quad 
\epem\to Z^*\to \cnone\cnone,\cnone\cntwo,\cntwo\cntwo\,.
\end{equation}
If $|\delgs|$ is not too large,
the $\cnone$ and $\cntwo$ are primarily wino and bino, respectively,
with weak couplings to the $Z$, and
the latter neutralino pair cross sections are always much
smaller than the $\cpone\cmone$ cross section.\footnote{Our cross 
section results include
slepton and sneutrino exchanges in the $t$- and $u$-channels;
see Ref.~\cite{bartl} for explicit expressions. However,
these diagrams are suppressed for a $\rts\leq 500\gev$ collider
by the large selectron and sneutrino
masses deriving from the large magnitude of $m_0$ (when $|\delgs|$
is not extremely large).}
As $\rts$ increases, 
\begin{equation}
\epem\to \cpmone\cmptwo\,,\quad \epem\to \chitil^0_{1,2}
\chitil^0_{3,4}
\end{equation}
become kinematically allowed. When allowed, the latter gaugino-higgsino 
(light-heavy) neutralino
pair cross sections are larger than the gaugino-gaugino 
(light-light) neutralino
pair cross sections due to the large higgsino components of the heavy
neutralinos. At still higher $\rts$, typically above (below) $500\gev$
if $\tanb$ is small (large), the
\begin{equation}
\epem\to \cptwo\cmtwo\,,\quad\epem\to \chitil^0_{3,4}\chitil^0_{3,4}\,,
\end{equation}
processes become possible. When allowed,
the higgsino-higgsino (heavy-heavy) 
neutralino pair cross sections are comparable
to chargino pair cross sections. 

As $-\delgs$ increases in magnitude, the bino/wino content 
of the $\cnone$ and $\cntwo$ becomes more mixed, but the general cross
section expectations are not greatly altered since $\mu$ is
always sufficiently large that it is the $\cptwo$, $\cnthree$
and $\cnfour$ which remain primarily higgsino.

\begin{figure}[htb]
\begin{center}
\centerline{\psfig{file=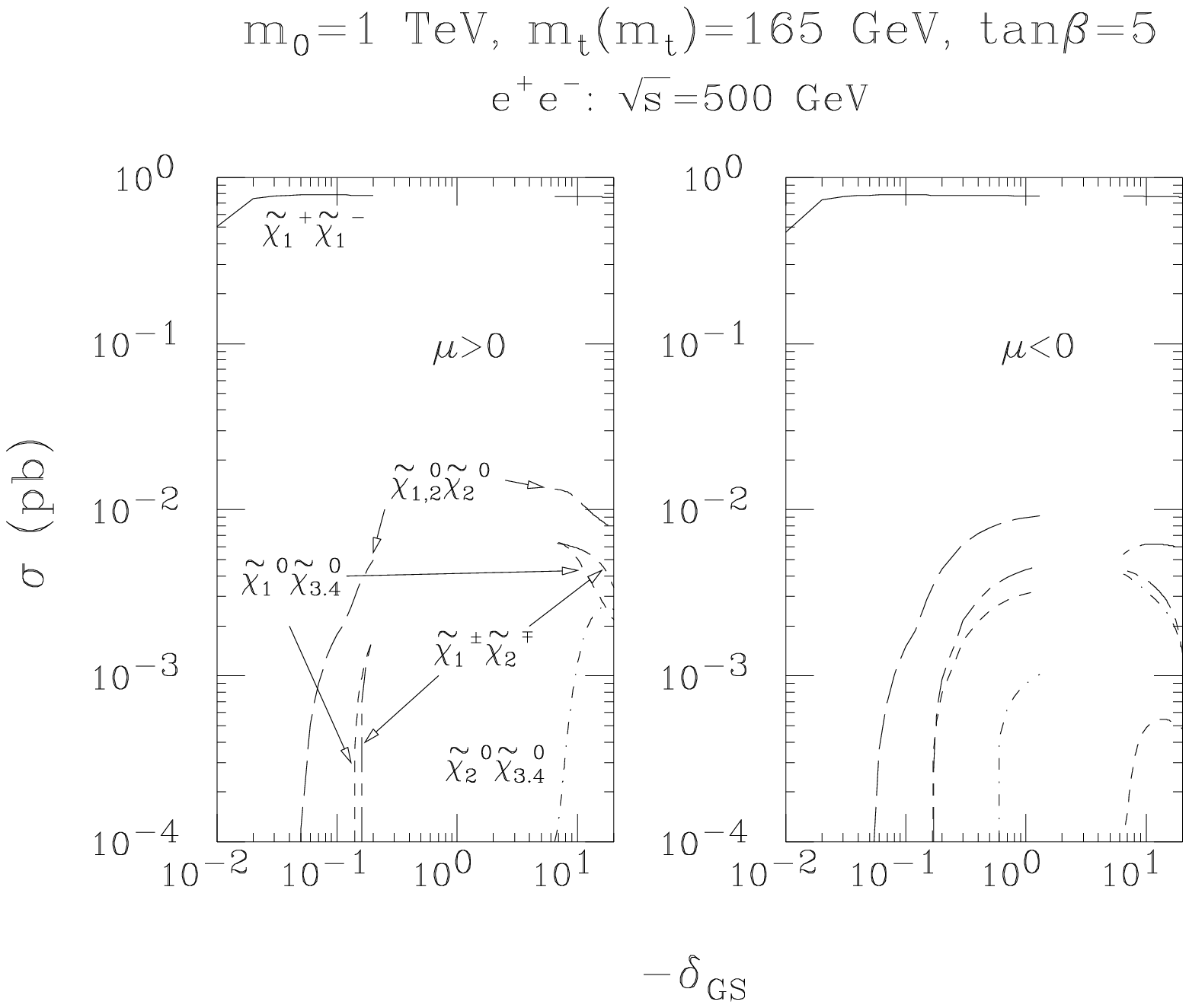,width=11cm}}
\bigskip
\begin{minipage}{12.5cm}       
\caption{Neutralino and chargino pair cross sections in $\epem$ collisions at 
$\protect\rts=500\gev$
as a function of $-\delgs$, taking $m_0=1\tev$ and $\tanb=5$.
Results are displayed for both signs of $\mu$.  The gap in $-\delgs$
is where $\mgl\leq 120\gev$ ($\mu<0$), 
a very rough limit from Tevatron data, or $\mcpmone<45\gev$ ($\mu>0$),
as excluded by LEP1.
The legend is: solid -- $\cpone\cmone$; 
long dashes -- $\cntwo\chitil_{1,2}^0$;
short dashes -- $\cnone\chitil_{3,4}^0$;
dotdash -- $\cntwo\chitil_{3,4}^0$;
long dash/short dash -- $\cpmone\cmptwo$.
}
\label{sigdgnp}
\end{minipage}
\end{center}
\end{figure}

\begin{figure}[htb]
\begin{center}
\centerline{\psfig{file=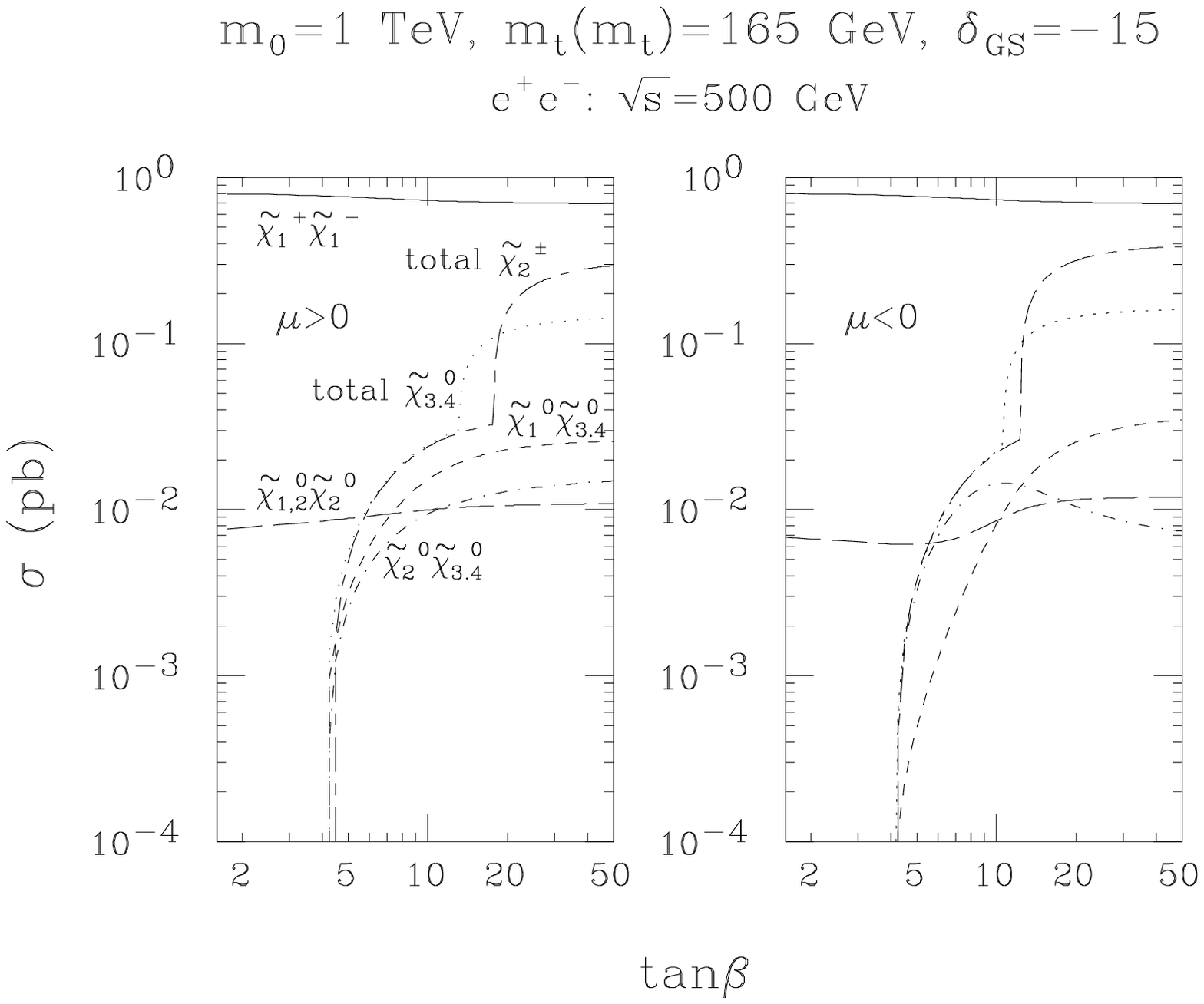,width=12cm}}
\bigskip
\begin{minipage}{12.5cm}       
\caption{Neutralino and chargino pair cross sections in $\epem$ collisions at 
$\protect\rts=500\gev$
as a function of $\tanb$, taking $m_0=1\tev$ and $\delgs=-15$.
Results are displayed for both signs of $\mu$.  
The legend is: solid -- $\cpone\cmone$; 
long dashes -- $\cntwo\chitil_{1,2}^0$;
short dashes -- $\cnone\chitil_{3,4}^0$;
dotdash -- $\cntwo\chitil_{3,4}^0$;
long dash/short dash -- $\cpmone\cmptwo+\cpmtwo\cmptwo$;
dots -- $\cnone\chitil_{3,4}^0+\cntwo\chitil_{3,4}^0
+\chitil_{3,4}^0\chitil_{3,4}^0$
}
\label{sigtanbnp}
\end{minipage}
\end{center}
\end{figure}

The relevant cross sections are illustrated for $\rts=500\gev$
in Figs.~\ref{sigdgnp} and \ref{sigtanbnp}. 
In Fig.~\ref{sigdgnp} the gaps in $-\delgs$ are where
$\mgl$ falls below $120\gev$, which includes
the region where the gluino would be the LSP. The $120\gev$ 
lower limit is a rough 
characterization of the bound from Tevatron data in this model.
A detailed examination of Tevatron predictions in a later section
shows that the actual bound varies significantly as a function of
$-\delgs$.  For example,
values of $\mgl$ below $100\gev$ are still allowed by current analyses
if $\mgl\sim \mchi$, as happens in two narrow bands within
the gap region (see Fig.~\ref{inos}).

As anticipated, the $\cpone\cmone$ cross
section is far and away the largest, but for $-\delgs<7-10$
can be quite difficult to see by virtue of the softness of
the $\cpmone$ decay products.  However, the $\cntwo\chitil^0_{1,2}$
cross section and the various gaugino-higgsino cross sections
can be kinematically allowed and large enough to be
observable.  For all these latter processes 
the final state should contain
some energetic leptons/jets in association with a
large amount of missing energy and should be easily detected
if the event rate is adequate.  For the target yearly luminosity
of $L=50\fbi$ at $\rts=500\gev$, one can probably be sensitive
to a raw cross section as small as $10^{-3}\pb$ (yielding 50 events
before cuts) in such final states.
Fig.~\ref{sigdgnp} shows that if $-\delgs$ is not below $\sim 0.05$,
then at least one of these visible final states will have adequate
cross section. 

The first visible final states to become accessible
as $-\delgs$ increases are $\cnone\cntwo$ and $\cntwo\cntwo$.
These are not kinematically allowed at small $-\delgs$ since 
$\mcntwo$  becomes
a factor of three larger than $\mchi$,
and is growing $\propto 1/\sqrt{-\delgs}$ [see Eq.~(\ref{smalld})].
The next channels to open up as $-\delgs$ increases are
$\cnone\chitil_{3,4}^0$ and $\cpmone\cmptwo$.  The thresholds
for these two final states are very similar due to the
degeneracies $\mcnone\sim\mcpmone$ and $\mcnthree\sim\mcnfour\sim\mcpmtwo$.
The cross sections are of limited magnitude because they are
only non-zero to the extent that there is some higgsino-gaugino mixing
in the mass eigenstate compositions. As $\tanb$ increases,
Fig.~\ref{par} shows that $|\mu|$ decreases substantially,
implying that the higgsino states become lighter (see Fig.~\ref{inos}).
This also implies greater gaugino-higgsino mixing. Thus, as seen in
Fig.~\ref{sigtanbnp}, there is a $\tanb$ threshold in the
$\tanb\gsim 10-15$ region where the total $\cpmtwo$ and total
$\cnthree,\cnfour$ cross sections suddenly increase due to the fact
that the higgsino states become light enough 
that they can also be pair produced at $\rts=500\gev$.
Since the $Z\to$higgsino+higgsino coupling is large, the higgsino
pair cross sections are comparable to the chargino-pair cross section.
Although the $\cnthree$ and $\cnfour$ are primarily higgsino
in nature, their SU(2) gaugino content 
is generally sufficient that their dominant
decay is to $\wpm\cmpone$, rather than to 
$\hl\cnone,\hl\cntwo$ (where $\hl$ is the light CP-even Higgs boson)
that would dominate if they were pure higgsino.

\begin{figure}[htb]
\begin{center}
\centerline{\psfig{file=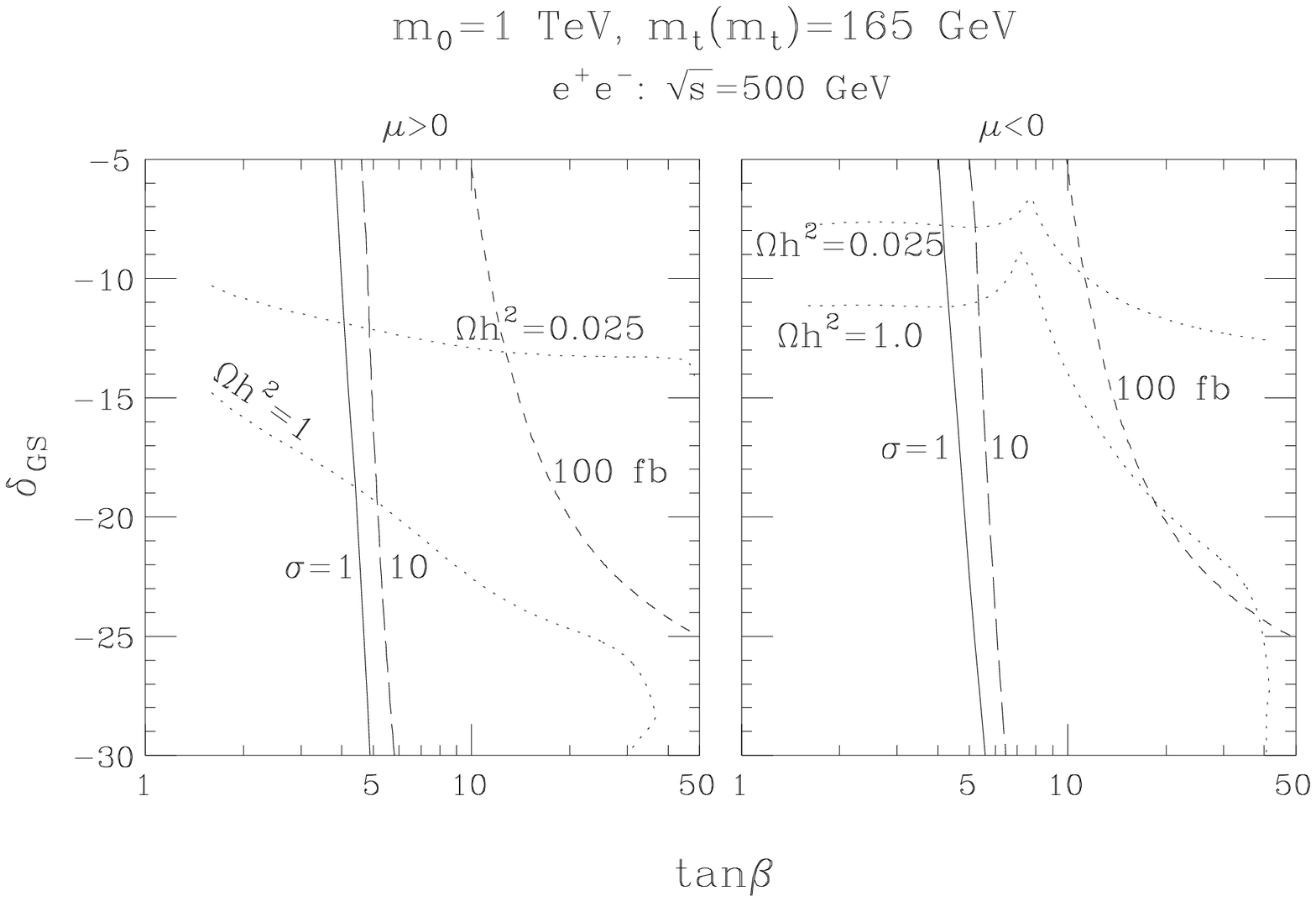,width=12cm}}
\bigskip
\begin{minipage}{12.5cm}       
\caption{Contours of constant $\sigma$ [Eq.~(\protect\ref{sigdef})] in
$\tanb$--$\delgs$ parameter space for $m_0=1\tev$ and $\protect\rts=500\gev$.
Also shown (dots) are the $\Omega h^2=1$ and $0.025$ contours ---
see Section 2.5}
\label{sigcontours}
\end{minipage}
\end{center}
\end{figure}

In Fig.~\ref{sigcontours}, we present contours of 
\begin{equation}
\sigma\equiv \sum_i \left[\sigma(\chitil_i^0\cnthree)
+\sigma(\chitil_i^0\cnfour)+\sigma(\chitil_i^\pm\cmptwo)\right]
\label{sigdef}
\end{equation}
in the $\tanb$--$\delgs$ parameter plane for a $\rts=500\gev$
$\epem$ collider. Detection of at least one of the heavier
neutralino or chargino states is important as a means
for determining $|\mu|$ (from the fact that
the states typically have mass $\sim |\mu|$), thereby allowing
a check of the consistency of electroweak symmetry breaking.
Figure~\ref{sigcontours} focuses on the $-\delgs\geq 5$ domain.
As previously noted, for increasing $\tanb$ the value of
$|\mu|$ declines and the higgsino masses
decrease. Thus, $\sigma$ will first become non-zero when
gaugino+higgsino production becomes possible, with a second
very rapid increase as one crosses the higgsino+higgsino threshold.
Fig.~\ref{sigcontours} shows that the gaugino+higgsino threshold
lies at $\tanb\sim 5$ for $m_0=1\tev$.  
For $\tanb\gsim 5$, $\sigma\gsim 1\fb$
(implying $\gsim 50 $ events) and 
detection of the heavy inos should be possible.
As noted earlier, $\cnthree,\cnfour$ will tend to decay to
$\wpm\cmpone$. The $\cpmtwo$ will decay primarily to $Z\cpmone$ 
or $\wpm\cnone,\wpm\cntwo$.
Since the thresholds for the gaugino+higgsino final states
are very steep, $\sigma$ rises to $10\fb$, implying
500 events, already by $\tanb\sim 6-7$.  However,
since $\sigma$ is the sum over a number of modes, 500
events might still not be enough to make a precise 
determination of the masses of the $\cpmtwo$,
$\cnthree$, and $\cnfour$ and other model parameters
to which the cross sections are sensitive. For this, $\sigma\sim 100\fb$
(5000 events divided up among the channels) might be required.
This level of cross section is generally only achieved when 
higgsino+higgsino pair production is possible, which typically
requires fairly large $\tanb$. The $\sigma=100\fb$ contour
in Fig.~\ref{sigcontours} is close to the higgsino+higgsino threshold.

\begin{figure}[htb]
\begin{center}
\centerline{\psfig{file=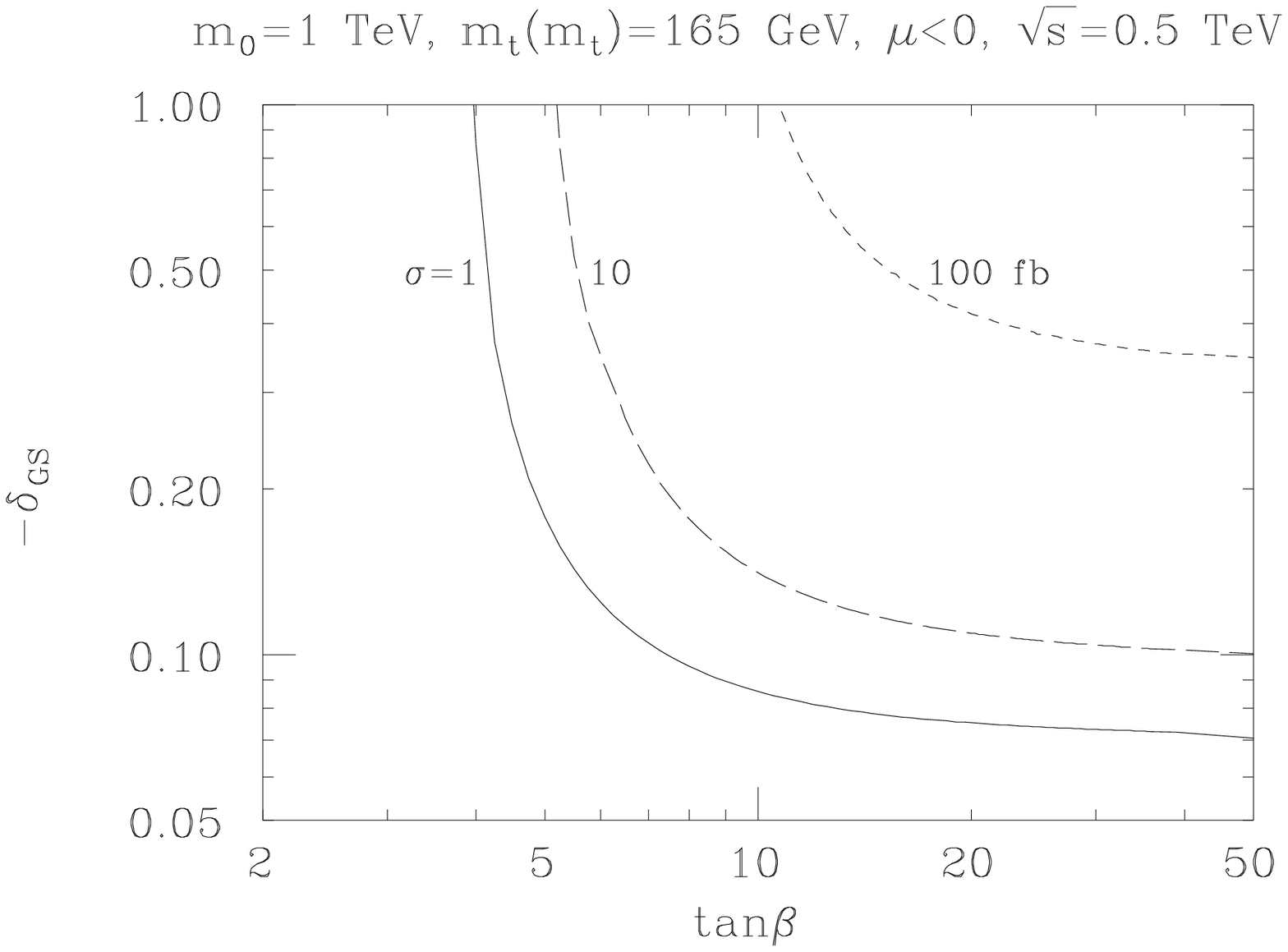,width=12cm}}
\bigskip
\begin{minipage}{12.5cm}       
\caption{Contours of constant $\sigma$ [Eq.~(\ref{sigdef})] in
the small $-\delgs$ portion of 
$\tanb$--$\delgs$ parameter space for $m_0=1\tev$ and $\protect\rts=500\gev$.
}
\label{sigcontourssmalldgs}
\end{minipage}
\end{center}
\end{figure}

These same $\sigma$ contours are displayed for small $-\delgs$
in Fig.~\ref{sigcontourssmalldgs}. As for larger $-\delgs$,
the $\sigma=1\fb$ contour is more or less defined by the
onset of gaugino+higgsino production, and the $\sigma=100\fb$
contour by the higgsino+higgsino threshold.  Note that
at small $-\delgs$ the masses of all the inos become large and 
ino pair production is not allowed for $\rts=500\gev$ if $m_0=1\tev$.
However, if $-\delgs$ is small then $m_0$, which sets the scale
for all masses in this model, can be lowered substantially
before $\mcpmone$ falls below the LEP limit of $47\gev$.

\subsubsection{Using photon tagging to detect {\boldmath $\cpmone\cmpone$}
pair production}

The most delicate question is whether $\cpone\cmone$ production
is observable when $\dmchi$ is small. As discussed
earlier (Fig.~\ref{lifebrs} and associated text),
for $m_\pi<\dmchi\lsim 1\gev$ the $c\tau$ of the $\cpmone$ will
be such that either the heavily-ionizing track of the $\cpmone$
or the non-zero impact parameter of the single charged pion
it emits is likely to be visible in an NLC vertex detector. 
[Since the innermost layer of the vertex detectors of the LEP
experiments is much further out ($r\sim 6.3$ cm) than at the NLC
($r\sim 1-2$ cm), this will be difficult at LEP.]
For $\dmchi<m_\pi$ the
$\cpmone$ will have a path length of order meters and will be
visible as a heavily-ionizing track in both the vertex detector and
the tracker. Events with two such tracks would provide a clear signal.
For large enough $\dmchi$, the leptons from the $\cpone$ and $\cmone$ decays
become visible as their momenta spectra
extend out beyond $p_T^{\ell}\gsim 1\gev$; the required
$\dmchi$ depends upon $\mcpmone$ and $\rts$.
Still, it is problematical that events with such soft leptons
could be isolated from two-photon backgrounds and the like.
In particular, unless $\dmchi>2\gev$, the final states
arising in $\epem\to\cpone\cmone$  production and decay are
similar (\ie\ contain leptons and missing energy)
to those appearing in $\gam\gam\to \tau^+\tau^-$ production and decay.
This latter background will be very large and
difficult to overcome, even if the chargino pair events can be triggered on.
There are ongoing analyzes by the LEP experimental groups
of their sensitivity to $\cpone\cmone$ production when $\dmchi$ is small.
Similar analyzes at NLC energies are also needed.  There,
the leptons are somewhat harder for given values of $\dmchi$ and $\mchi$
because of the higher energy, but the detector has a larger
magnetic field (designed to curl up the soft leptons from beamstrahlung
and related sources). Because of the detailed level of experimental
simulation required to address these questions, we will not
pursue detection of $\epem\to\cpone\cmone$ production at small $\dmchi$
further in this paper.

In Ref.~\cite{cdgi} we examined $\epem\to \gam\cpone\cmone$
production to see if a signal could
be observed above background at LEP-II and the NLC.
Here a hard-photon tag provides a trigger for the presence
of the $\cpone\cmone$ pair. 
We found that the range of $\mcpmone$ accessible via this final state 
depends greatly upon whether the $\cpmone$ decays are in any way
visible. This is because $\epem\to \gam\nu\anti\nu$ (via $\gam Z^*$)
becomes a very large background if the $\cpmone$ decay invisibly. 
From Fig.~\ref{lifebrs} we observe that for $\dmchi<m_\pi$, the $\cpmone$
are long-lived and create easily observed heavily-ionizing
tracks. The photon tag is probably not even needed.
As discussed earlier, 
for $m_\pi<\dmchi<1\gev$, the $\cpmone\to \pi^{\pm}\cnone$ decay mode 
is dominant and the $c\tau$ is such that either the 
heavily ionizing track of the $\cpmone$ or the non-zero impact
parameter of the $\pi^\pm$ it emits will be visible using 
the NLC vertex detector, but probably not using the LEP vertex detectors.
For $\dmchi>0.5\gev$, the $\cpmone\to \ell^{\pm}\nu\cnone$ mode
has branching ratio $>10\%$ for $\ell=e,\mu$ and one can look
for soft leptons. Indeed, for $\dmchi>1\gev$, the $\cpmone$ decay becomes
prompt and this would be the only signal for the decay.
However, detection of the soft charged pion or soft lepton
from a $\cpmone$ decay might be difficult since
soft charged particles may not reach the calorimeter in the
presence of a magnetic field. [Note that since the average energy
of these soft charged particles increases in going from LEP energy
to NLC energy (keeping $\mcpmone$ and $\dmchi$ fixed) it
is not clear that this difficulty will be
more severe at the NLC than at LEP, despite the previously mentioned
higher magnetic field of the NLC detector.]

\begin{figure}[htb]
\begin{center}
\centerline{\psfig{file=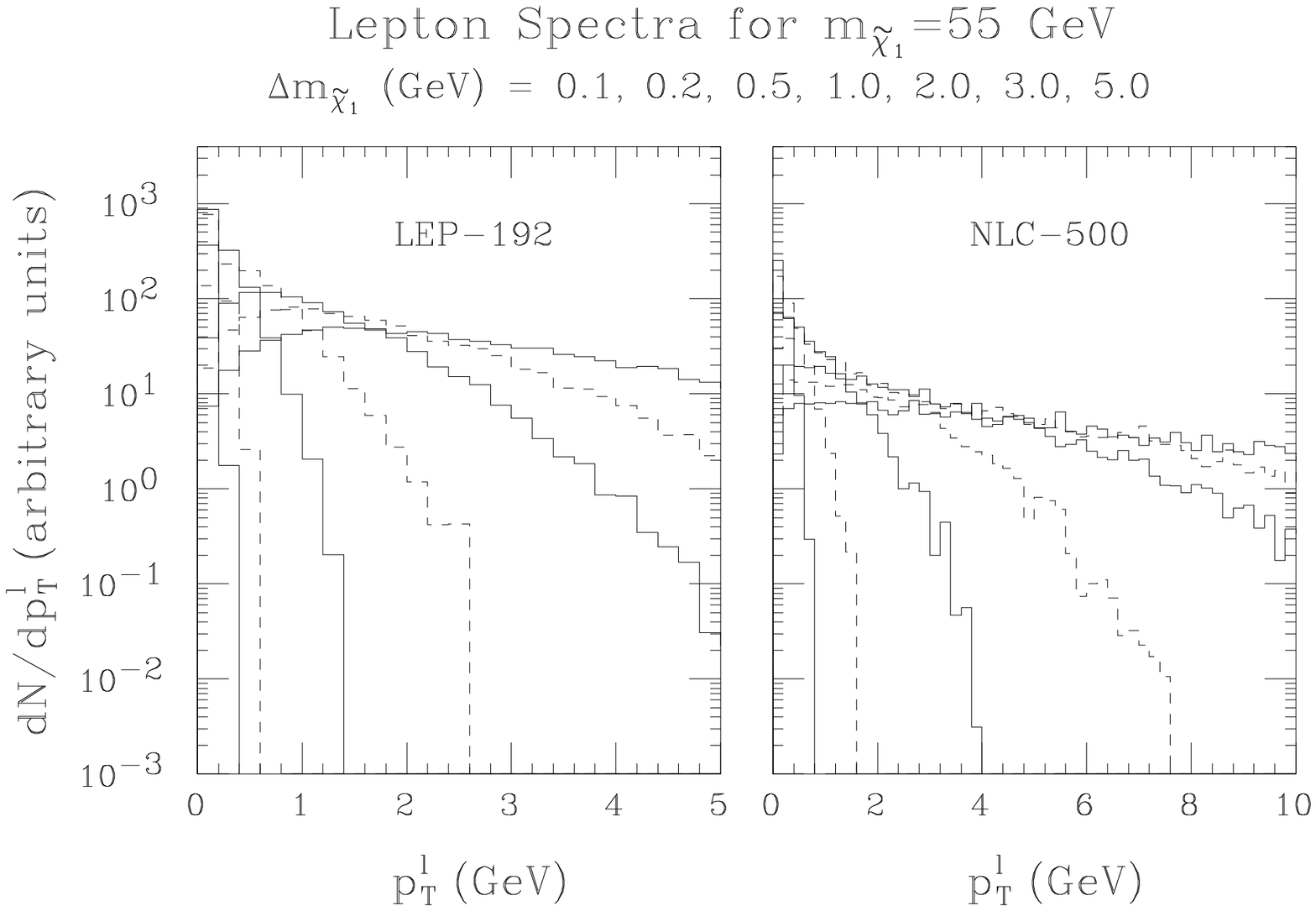,width=12cm}}
\bigskip
\begin{minipage}{12.5cm}       
\caption{$dN/dp_T^\ell$ vs. $p_T^\ell$ for the soft leptons in
$\epem\to\gam\cpone\cmone$ followed by $\cpmone\to\ell\nu\cnone$
in arbitrary units, taking
$\mchi\equiv\mcpmone=55\gev$ and $\dmchi\equiv \mcpmone-\mcnone$
values of 0.1, 0.2, 0.5, 1, 2, 3, and 5 GeV.  Results are given
for LEP operating at $\protect\rts=192\gev$ and NLC at $\protect\rts=500\gev$.
}
\label{spectra}
\end{minipage}
\end{center}
\end{figure}

In order to provide a more quantitative picture of the difficulty
of seeing the soft leptons coming from the $\cpone\cmone$
pair, we present in Fig.~\ref{spectra} the $p_T^{\ell}$ spectra
deriving from $\cpmone\to\ell\nu\cnone$ decays in the $\gam\cpone\cmone$
final state for a variety of $\dmchi$ values, taking $\mcpmone=55\gev$.
We see, for example, that to observe leptons in a significant fraction of
the events when $\dmchi\sim 0.5\gev$, it is necessary to 
have good efficiency down to at least $0.7\gev$ ($2\gev$) at LEP-192
(NLC-500). These lepton spectra become still softer
as $\mcpmone$ increases keeping $\dmchi$ fixed. For example, at NLC-500
the lepton spectra for $\mcpmone=175\gev$ typically terminate
at about 1/2 the maximum value found for $\mcpmone=55\gev$,
keeping a fixed $\dmchi$ value.
It will be important for the experimental groups
to study how well they can do and if there are any detector changes
that might increase their sensitivity to soft leptons.

If the $\gam\nu\anti\nu$ background cannot be eliminated by
tagging the soft decay products or short tracks of the $\cpmone$, then
we must consider the best strategy for isolating the $\gam\cpone\cmone$
signal in its presence.  At the same time, we must be careful
to avoid additional backgrounds, the most dangerous being
that from $\epem\to\epem\gam$, where both the final $e^+$ and $e^-$
disappear down the beam hole. 

In the study of Ref.~\cite{cdgi},
we found a very effective procedure for eliminating
the $\epem\to\epem\gam$ background. We begin by
requiring a photon tag with  $p_T^\gam\geq \ptmin=10\gev$
and $10^\circ\leq\thetagam\leq 170^\circ$, where $\thetagam$
is the angle of the photon with respect to the beam axis. This
guarantees that the photon enters a typical detector and
will have an accurately measured momentum. We define $\gam+\etmiss$
events by requiring that any other particle appearing in the $10^\circ$
to $170^\circ$ angular range must have energy less than $\emax$,
where $\emax$ is detector-dependent, but presumably
no larger than a few GeV. Kinematics can be used to show that
we can then eliminate the $\epem\rta\epem\gam$ background
by vetoing events containing an $e^+$ or $e^-$ 
with $E>50\gev$ and angle $\thetamin\leq\thetae\leq 10^\circ$
with respect to either beam axis, or with $E>\emax$
and $10^\circ\leq\thetae\leq170^\circ$,
{\it provided} $\ptmin\gsim \sqrt s\sin\thetamin(1+\sin\thetamin)^{-1}$
(assuming $\emax$ is not larger than a few GeV). For $\ptmin=10\gev$,
this means that we must instrument the beam hole down to
$\thetamin=1.17^\circ$. In fact, instrumentation and vetoing
will be possible down to $\thetamin=0.72^\circ$ \cite{tbarklow},
implying that $\ptmin$ could be lowered to $\sim 6.2\gev$ without
contamination from the $\epem\rta\epem\gam$ background.
At LEP-192, beam hole coverage down to about $3.1^\circ$
is needed when using a $\ptmin=10\gev$ cut.

\begin{figure}[htb]
\begin{center}
\centerline{\psfig{file=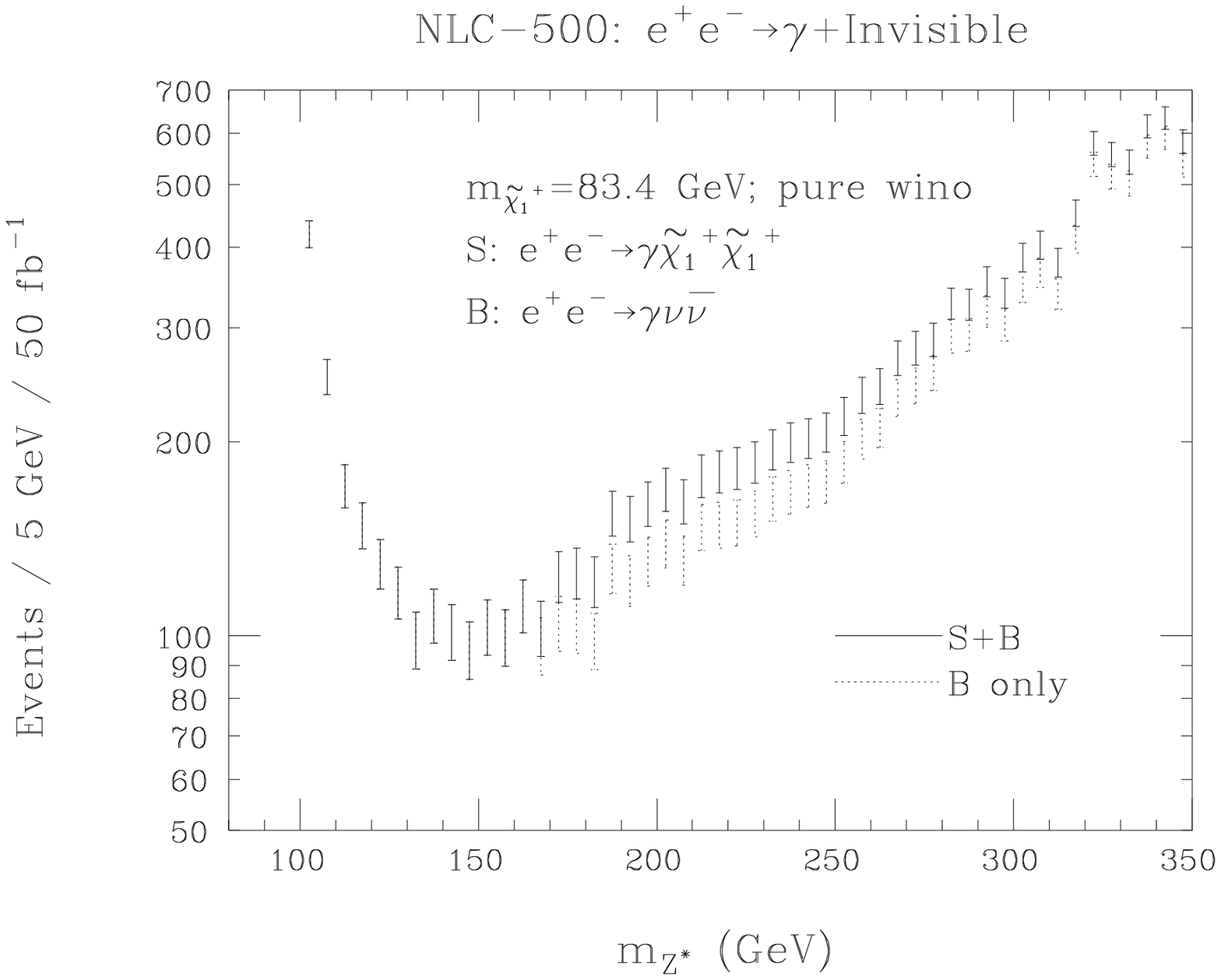,width=12cm}}
\bigskip
\begin{minipage}{12.5cm}       
\caption{For $\epem\to \gam+\etmiss$, 
we plot the the number of events per 5 GeV bin
per $L=50\fbi$ at NLC-500 as a function of $\mzstar$.
Solid error bars are for the sum of $\gam\cpone\cmone+\gam\nu\anti\nu$,
while dotted error bars indicate expectations for $\gam\nu\anti\nu$ alone.
We have chosen a scenario with $\mcpmone=83.4\gev$ in which the $\cpmone$
is pure wino.  Photon cuts are as described in the text.}
\label{mzstarchi1}
\end{minipage}
\end{center}
\end{figure}

For chargino masses above $\mz/2$, the key observation for reducing
the background from $\gam\nu\anti\nu$ and determining the chargino mass
is to note that the missing mass
$\mzstar\equiv [(p^{e^+}+p^{e^-} - p^\gam)^2]^{1/2}$ 
can be very accurately reconstructed.\footnote{We are
uncertain as to the extent to which beamstrahlung 
might impact our ability to compute the true $\zstar$ system mass.
Since most beamstrahlung
involves radiation of just one hard photon along the beam line,
the $\mzstar$ computed as above would correspond to the invariant mass
of the $\zstar+\gam_{\rm beamstrahlung}$ system, which is larger
than the mass of the $\zstar$ alone. The seriousness of this
effect will depend on the machine parameters, as well as on the chargino
mass. For heavier charginos the cut on $\mzstar$ becomes
stronger, so contamination from background events
with hard collinear photons becomes less likely. Machine parameters
for which the beamstrahlung photon typically carries less than 10\%
of the beam energy should not greatly distort $\mzstar$; our
ability to make the $\mzstar$ cut and measure the
threshold onset of $\gam\cpone\cmone$ would not be significantly impaired.} 
For signals with
good overall statistical significance (in most cases $\nsd$, defined below,
$\gsim 5$ is adequate) one can plot events
as a function of $\mzstar$ and look for the threshold at $2\mchi$
at which the spectrum starts to exceed the expectations from
$\gam\nu\anti\nu$.  This is illustrated for NLC-500 in Fig.~\ref{mzstarchi1},
where we plot the number of events per 5 GeV bin as a function of
$\mzstar$ assuming $L=50\fbi$, comparing a $\mchi=83.4\gev$
signal with expectations for the $\gam\nu\anti\nu$ background alone.
Expected error bars are shown.  
We see that a determination of the threshold within
about $\pm 10$ to $\pm15\gev$ should prove possible in this case.

\begin{figure}[htb]
\begin{center}
\centerline{\psfig{file=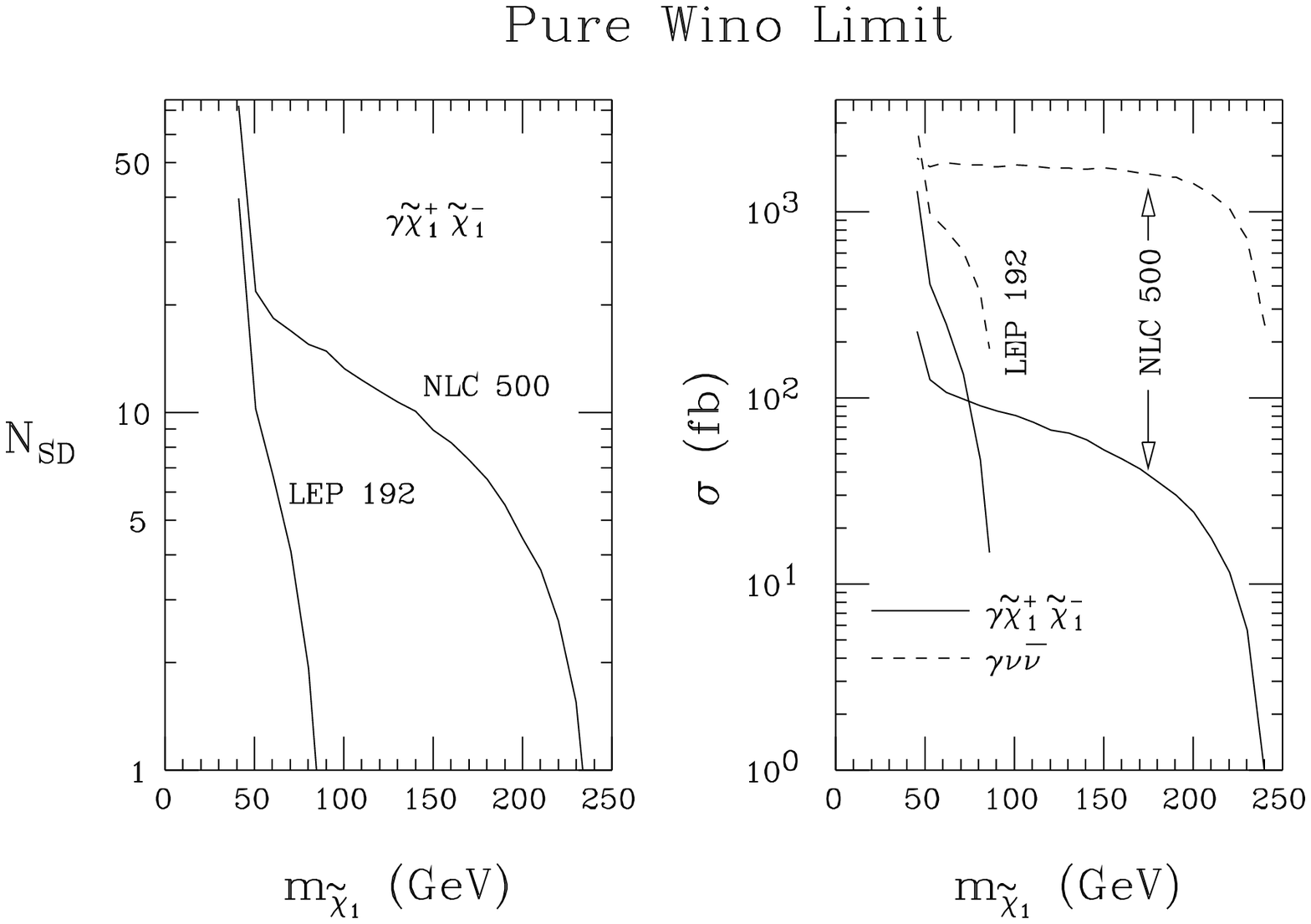,width=12cm}}
\bigskip
\begin{minipage}{12.5cm}       
\caption{We plot the statistical significance $\nsd=S/\protect\sqrt B$
for detecting $\gam\cpone\cmone$ in the $\gam+\etmiss$ channel 
as a function of $\mcpmone$. The background rate, $B$, is computed from
$\epem\rta\gam\nu\anti\nu$ by integrating over $\mzstar\geq 2\mcpmone$.
Results for LEP-192 (with $L=0.5\fbi$) and NLC-500 (with $L=50\fbi$)
are displayed. Also shown are the $\gam\cpone\cmone$
and $\gam\nu\anti\nu$ cross sections themselves.
We employ the cuts $p_T^\gam\geq 10\gev$ and
$10^\circ\leq\thetagam\leq170^\circ$. Results are for a chargino
that is pure wino; slepton and sneutrino masses are assumed to be large.
}
\label{eetogaminv}
\end{minipage}
\end{center}
\end{figure}

We define an overall statistical significance 
$\nsd=S/\sqrt B$ for the signal by summing over all events with 
$\mzstar>2\mcpmone$. Note, in particular, that
this cut eliminates the $Z$-pole contribution to the $\gam\nu\anti\nu$
background when $\mcpmone>\mz/2$. 
The results for $\nsd$ as a function of $\mcpmone$,
as well as the $S=\gam\cpone\cmone$ and $B=\gam\nu\anti\nu$ 
cross sections (after integrating
over $\mzstar\geq 2\mcpmone$), are plotted in Fig.~\ref{eetogaminv}.
For the particular example
of Fig.~\ref{mzstarchi1} ($\mcpmone=83.4\gev$, $\rts=500\gev$
and $L=50\fbi$), we find $S/\sqrt B\sim 15$.
In practice, one can often do better (perhaps by $1\sigma$ to $2\sigma$)
than the nominal $\nsd$ values plotted
in Fig.~\ref{eetogaminv} by zeroing in on those $\mzstar$
bins with the largest deviations from $\gam\nu\anti\nu$ expectations.

From the results of Fig.~\ref{eetogaminv} we see that 
at NLC-500 (LEP-192) $\nsd=5$
is achieved for $\mcpmone\lsim 200\gev$ ($\lsim 65\gev$).  Thus, one could
not probe all the way to the $\mcpmone\sim \rts/2$ kinematical limit,
as would be possible in the $\epem\to\cpone\cmone$ channel for conventional
universal boundary conditions. 

The situation is very different if one can detect the (soft) decay
products or vertex tracks 
of the $\cpone\cmone$ in the $\gam\cpone\cmone$ final state,
since then $\gam\nu\anti\nu$ production is no longer a background.
The only background requiring discussion is the background
from $\epem\to\epem\gam\ell^+\ell^-$, with two 
leptons disappearing down the beam pipe. This is a potential background
to final states in which both the $\cpone$ and $\cmone$
decay leptonically. However, it will be very small.
First, it will be greatly suppressed (although not entirely eliminated)
by using the vetoing procedure outlined above. 
Second, it is ${\cal O}(\alpha^5)$ vs. ${\cal O}(\alpha^3)$ for the signal.
Third, the enhancement deriving from singular $t$-channel photon exchange
is only operative and escapes the veto if the energetic $\epem$ both disappear
down the beam pipe, implying that the observed soft leptons 
(\ie\ the $\ell^+\ell^-$ pair) must be of the
same type; the background due to this configuration
could thus be eliminated by focusing on the soft $e^+\mu^-$ and
$\mu^+e^-$  pairs that are just as probable as $\mupmum$ or $\epem$ pairs
in $\cpone\cmone$ decays (assuming that we can distinguish a muon
from an electron at low energy). Only $\ell^+\ell^-=\tau^+\tau^-$ 
could yield $e\mu$ final states. Thus, we believe that backgrounds
to $\gam\cpone\cmone$ production are negligible when the soft $\cpone\cmone$
decay products are visible.

In the absence of significant background, the observability of
$\epem\to \gam\cpone\cmone$ followed by detection of the soft 
decay products or short tracks of the $\cpone\cmone$
depends entirely on event rate. The latter is simply given by 
the luminosity times the cross section plotted in Fig.~\ref{eetogaminv}.
Assuming 50 (background free) events are required, we could
detect $\gam\cpone\cmone$ at NLC-500 (LEP-192) all the way up
to $240\gev$ ($75\gev$) for $L=50\fbi$ ($0.5\fbi$). The 
increase in discovery range compared to the case where
the $\cpone\cmone$ are invisible to the detector is especially marked
at the NLC, with the mass reach improving
almost to the $\rts/2$ kinematical limit.

\subsection{SUSY Discovery at Hadron Colliders}

As noted earlier, detection of a signal from supersymmetric
particle production at a hadron collider need not be straightforward
in the O-II model scenario.  Since the squarks and sleptons
are necessarily very heavy in the O-II model (unless $|\delgs|$ is
very large), gluino-pair and electroweak-gaugino-pair production
would appear to provide the greatest potential for SUSY discovery.
While this is certainly the case at the Tevatron, gluino-squark
and squark-squark pair production at the LHC would be possible.
We will comment on these SUSY signals later.

First, we focus on electroweak-gaugino-pair and gluino-pair production.
The primary channel for detecting the former
is normally the $3\ell$ channel deriving from
$\cpmone\cntwo\to\ell^\pm\ell^+\ell^-\etmiss X$. If
$\dmchi$ is small (small $|\delgs|$) the $\ell$ from 
$\cpmone\to\ell^\pm\nu\cnone$ decay is so soft that
the $3\ell$ signal is negligible.
The two primary modes for detecting $\gl\gl$ pair production
are the $jets+\etmiss$ channel and the like-sign dilepton,
$\ell^\pm\ell^\pm+jets+\etmiss$ signal.
The like-sign dilepton signal
(from $\gl\gl\to \cpmone\cpmone X\to \ell^{\pm}\ell^{\pm}\etmiss X$) 
will be negligible if $\dmchi$ is small because of the
softness of the leptons. To the extent that it is not observable,
it will add to the $jets+\etmiss$ signal (which
is defined by events having no observable hard lepton).

\begin{figure}[htb]
\begin{center}
\centerline{\psfig{file=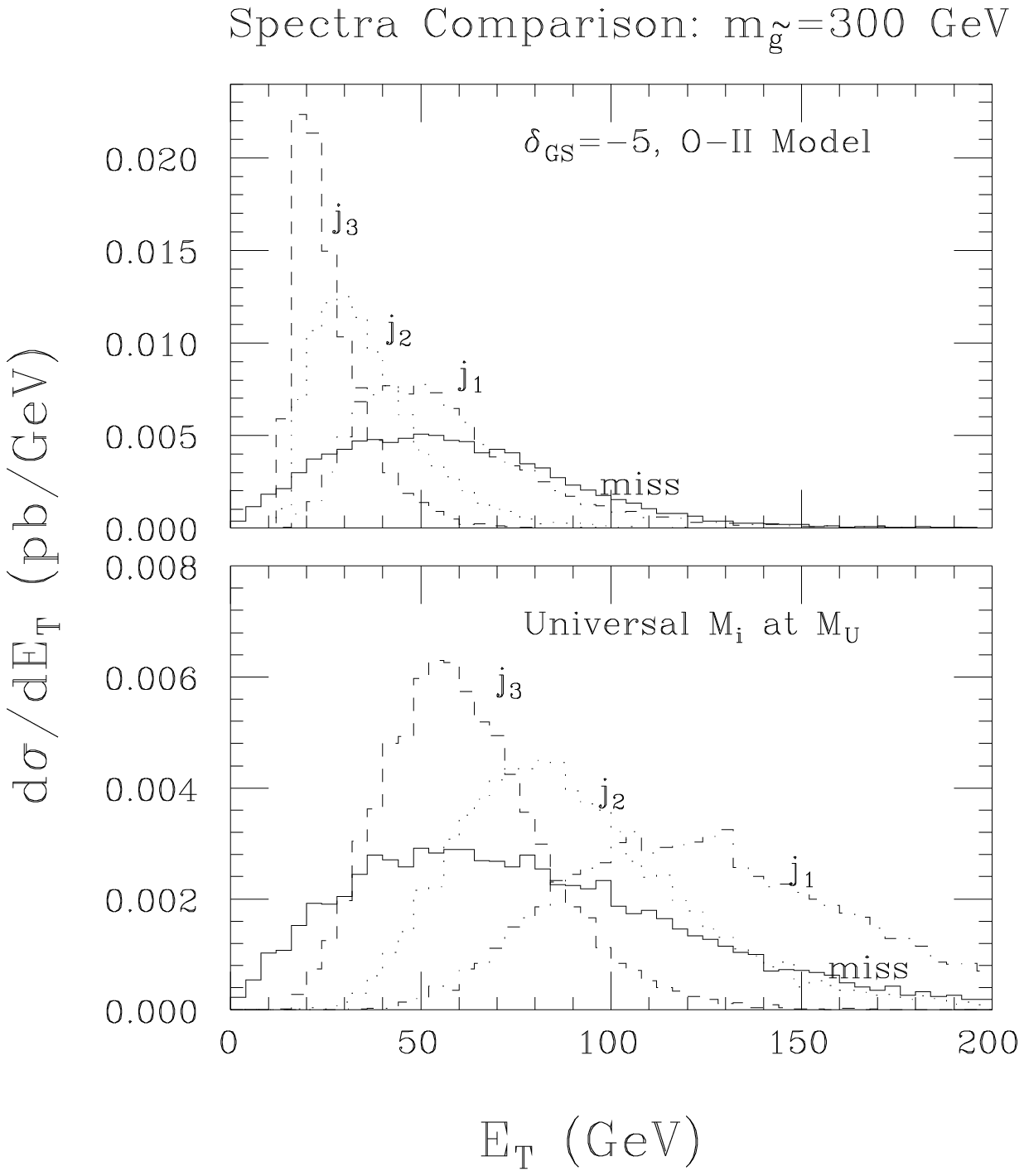,width=9.5cm}}
\bigskip
\begin{minipage}{12.5cm}       
\caption{Spectra, $d\sigma/dE_T$, versus $E_T$ for 
the three most energetic jets (labelled $j_1$, $j_2$, $j_3$
in order of decreasing $E_T$) and the missing energy.
The spectra for the universal boundary condition model and the O-II model
with $\delgs=-5$ are compared for $\mgl=300\gev$, $\tanb=2$
at the Tevatron. Jet-energy smearing effects are included.
}
\label{spectrumcompare}
\end{minipage}
\end{center}
\end{figure}

Depending upon the model parameters,
the $jets+\etmiss$ signal for $\gl\gl$ production may also
be difficult to isolate from background.
If $-\delgs$ is such that $\mgl\sim\mchi$,
the jets from $\gl$ decay are softer than for the universal
boundary condition models which have $\mgl\sim 3\mcpmone\sim 6\mcnone$.
This is illustrated in Fig.~\ref{spectrumcompare}.
There, the $E_T$ spectra for the most energetic
three jets and for $\etmiss$ as predicted
for universal boundary conditions and in the $\delgs=-5$ O-II
model are compared at the Tevatron. (This figure
includes the effects of jet energy smearing.) The jets are much harder
in the former case due to the large $\mgl-\mcpmone\sim200\gev$ 
and $\mgl-\mcnone\sim 250\gev$ mass
differences as compared to $\mgl-\mchi\sim 76\gev$
for the O-II model.  Correspondingly, $\etmiss$
is somewhat larger on average in the O-II model.
In the following, we determine
the portion of $\delgs$--$\mgl$ parameter space 
(equivalently $\delgs$--$m_0$ parameter space, see Fig.~\ref{mglcontours})
for which the $jets+\etmiss$ signal will be visible
at the Tevatron and Tev*.  

\subsubsection{The {\boldmath $jets+\etmiss$} SUSY Signal}

At the Tevatron or Tev*, we consider both D0 \cite{D0cuts}
and CDF \cite{CDFcuts}
cuts.  These are summarized below.\footnote{We presume
that the cuts of Ref.~\cite{D0cuts} that are
designed to eliminate jets formed around noisy
calorimeter cells and jets induced by particles from the main ring
accelerator do not significantly reduce the signal cross section.}

\underline{\bf D0 Cuts}
\begin{itemize}
\item There are no isolated leptons with $E_T>15\gev$, where
isolation is defined by requiring that additional $E_T$ within
$\Delta R\leq 0.3$ of the lepton be $<5\gev$. 
\item $\etmiss>75\gev$.
\item There are $n(jets)\geq 3$ jets having $|\eta_{\rm jet}|<3.5$
and $E_T>25\gev$, using a coalescence cone size of $\Delta R=0.5$.
These are ordered according to decreasing $E_T$ and labelled
by $k=1,2,3$.
\item The azimuthal separations of the $k=1,2,3$ jets
from the $\etmiss$ vector, $\delta\phi_k\equiv \Delta\phi(\etmiss,j_k)$,
are required to satisfy $0.1<\delta\phi_k<\pi-0.1$. It is further
required that
$\sqrt{(\delta\phi_1-\pi)^2+\delta\phi_2^2}>0.5$.
\end{itemize}
\underline{\bf CDF Cuts}
\begin{itemize}
\item There are no leptons with $E_T>10\gev$.
\item $\etmiss>60\gev$.
\item There are $n(jets)\geq 3$ jets having $|\eta_{\rm jet}|<2$
and $E_T>15\gev$, using a coalescence cone size of $\Delta R=0.5$.
\item Azimuthal separation requirements are the following:
$\Delta\phi(\etmiss,j_1)<160^\circ$; and
$\Delta\phi(\etmiss,j(E_T>20\gev))>30^\circ$. These are designed,
in particular, to reduce QCD jet mis-measurement.
\end{itemize}
We note that the lepton cut causes
no signal loss when the $\cpmone$ decay leptons
are very soft. (In comparison, in the
universal boundary condition scenarios signal events
sometimes have isolated leptons.)

Our procedure will be to compute the signal cross section, $\sigma_S$, at the
Tevatron as a function of $\delgs$ and $\mgl$ after imposing the
two different sets of cuts listed above.
For the background rates we take the D0 and CDF cross sections
as computed in Refs.~\cite{D0cuts,CDFcuts} for the above cuts, respectively.
For D0 cuts, the background cross section is taken from Table 1 of
Ref.~\cite{D0cuts} to be 16.7 events for $L=7.1\pbi$, corresponding
to a cross section of $\sigma_B=2.35\pb$. For CDF cuts, Ref.~\cite{CDFcuts}
quotes a background rate of 28.7 events for $L=19\pbi$, corresponding
to $\sigma_B=1.51\pb$. For a given luminosity, we compute the background
and signal rates as $S=L\sigma_S$ and $B=L\sigma_B$, respectively.
Both the D0 and CDF background computations
include full hadronic energy smearing and the like, so
that some of the background rate may come from fake $\etmiss$.
We also include the effects of hadronic energy smearing in the
signal rate computation.

\begin{figure}[htb]
\begin{center}
\centerline{\psfig{file=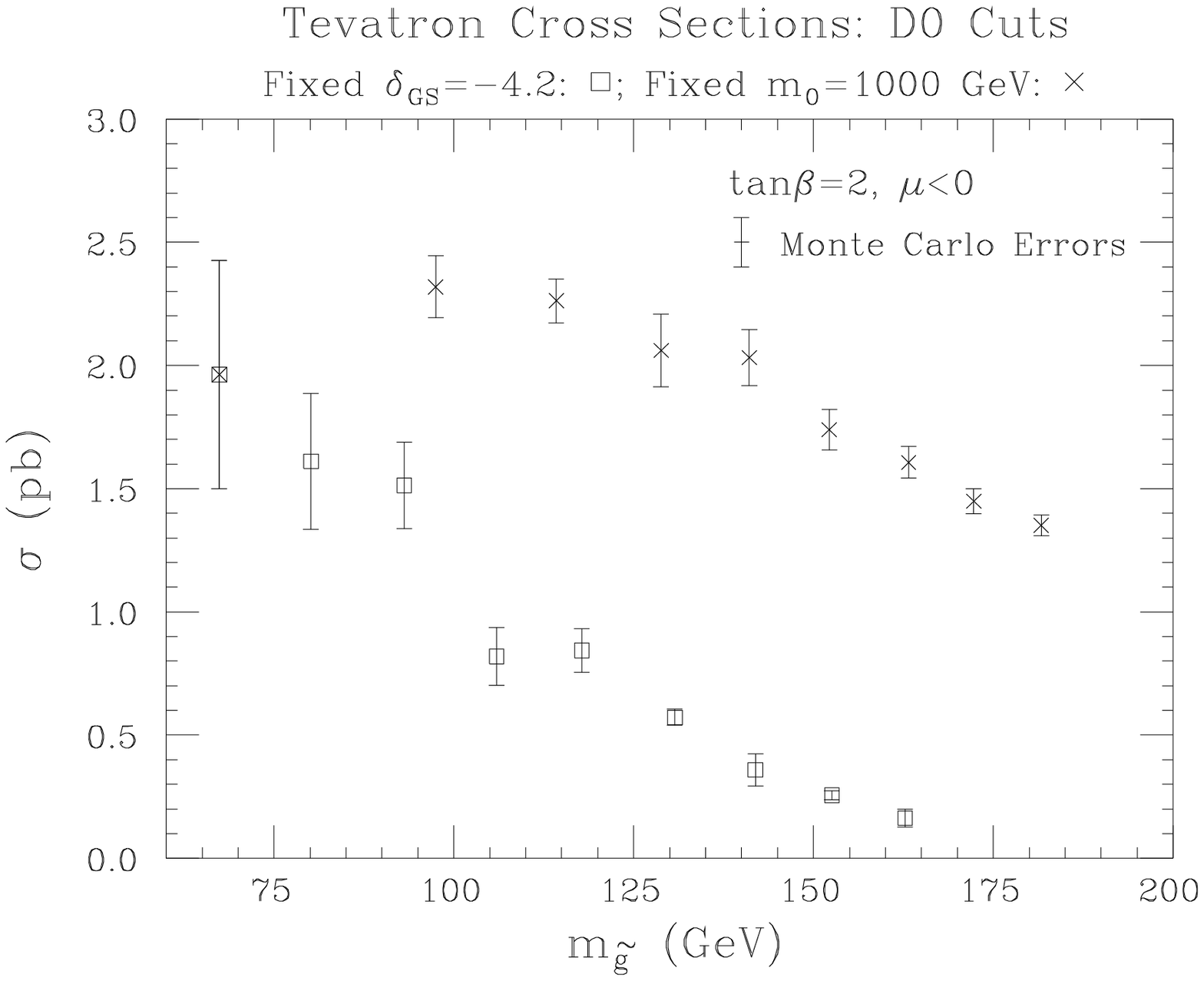,width=12cm}}
\bigskip
\begin{minipage}{12.5cm}       
\caption{Cross section after D0 cuts at the Tevatron as a function
of $\mgl$ for: (i) fixed $\delgs=-4.2$ along the
$\mgl\sim\mchi$ [see Eq.~(\ref{degdef})] boundary --- plotted points
correspond to $m_0=1$, $1.2$, $1.4$, $1.6$, $1.8$, $2.2$, $2.4$, and $2.6\tev$;
(ii) fixed $m_0=1\tev$ with increasing $-\delgs$ --- plotted points
are at $-\delgs=4.2$, 5, 5.5, 6, 6.5, 7, 7.5, 8, and 8.5.
We have taken $\tanb=2$ and $\mu<0$.}
\label{sigmavsmgluino}
\end{minipage}
\end{center}
\end{figure}

Before presenting an overall summary graph, it is useful to explicitly
demonstrate the impact of the softness of the jets when near
the $\mgl\sim\mchi$ boundary.  In Fig.~\ref{sigmavsmgluino}
the cross sections at the Tevatron, after imposing D0 cuts,
are displayed in two cases: 
\begin{itemize}
\item increasing $\mgl$ by increasing $m_0$ while holding fixed 
$\delgs=-4.2$ (\ie\ near the degeneracy boundary); and
\item increasing $\mgl$ by increasing $-\delgs$ while holding fixed
$m_0=1\tev$ (\ie\ moving rapidly away from the degeneracy boundary).
\end{itemize}
The rapid decline of the cross section in the former case, as
compared to the latter, is apparent.  The Tevatron will not
be able to detect $\gl\gl$ production out to as large an $\mgl$
along the degeneracy boundary as away from it. 

\begin{figure}[htb]
\begin{center}
\centerline{\psfig{file=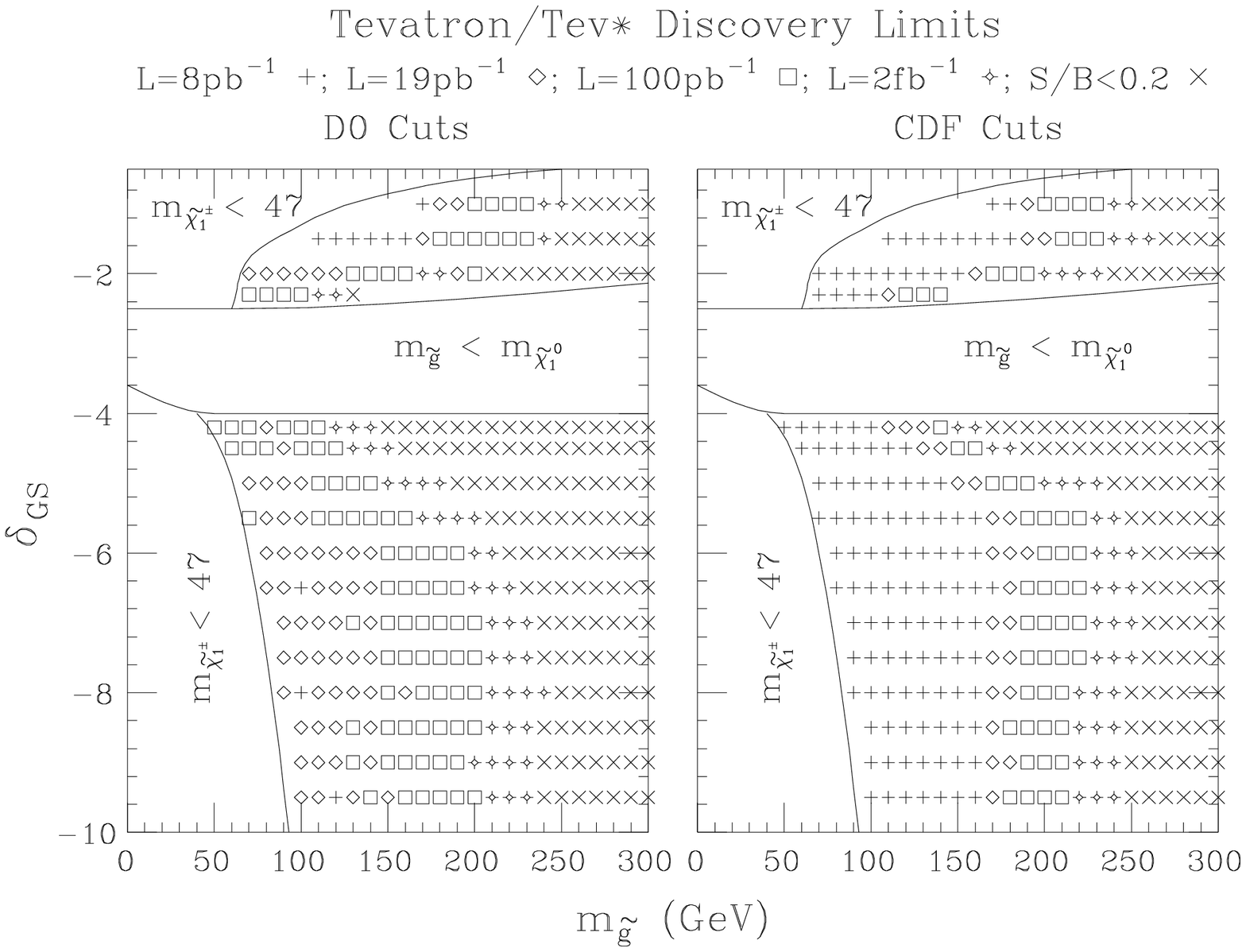,width=12cm}}
\bigskip
\begin{minipage}{12.5cm}       
\caption{O-II model regions of the $\delgs$--$\mgl$ parameter space
for which $\gl\gl$ production can be detected 
at the Tevatron/Tev* in the 
$jets+\etmiss$ channel for various different luminosities,
using the cuts described in the text.
Blank points are excluded, as indicated, either by the LEP
constraint of $\mcpmone\geq 47 GeV$ or by $\mgl<\mcnone$.
We have taken $\tanb=2$ and $\mu<0$.
}
\label{tevdiscovery}
\end{minipage}
\end{center}
\end{figure}

In Fig.~\ref{tevdiscovery} we show the $\delgs$--$\mgl$
parameter space regions for which the $jets+\etmiss$
signal should have been observed or will be observable for various
different integrated luminosities, when the D0 or CDF cuts outlined
above are employed.
Observability is defined by $S/\sqrt B\geq 5$ {\it and} $S/B\geq 0.2$,
where $S$ and $B$ are the numbers of signal and background events,
respectively, for a given luminosity.
The $S/B\geq 0.2$ requirement above is needed due to the
rather featureless nature of the signal which makes it
difficult to distinguish from the background using anything other
than the integrated cross section level. It is
the limiting factor in the maximum $\mgl$ value that
can be probed at high luminosity. The four different sets of
symbols indicate the following.
\begin{itemize}
\item Pluses indicate parameter space points that can be excluded by
analysis of roughly $L=8\pbi$ of data from Run-Ia. This is the amount
of data for which D0 has published an analysis and claimed
to see no signal.
\item Diamonds (together with pluses) 
indicate parameter space points excluded for $L=19\pbi$ 
of Run-Ia data. This is the amount of data
analyzed and published by CDF without observing a signal.
\item Squares indicate additional points that will be excluded for accumulated
luminosity of $L=100\pbi$, \ie\ if no signal is observed after
the full Run-Ia+Run-Ib data is analyzed.
Such analyzes should be available in the near future.
\item Next come the points indicated by a small star-like symbol
that can be excluded for $L=2\fbi$, \ie\ after one year
of running at the projected Main Injector luminosity.
\item For the points indicated by an $\times$, $S/B<0.2$ and
no amount of luminosity suffices.  Systematics would have to be
controlled at the $\lsim 10\%$ level to access this region.
\end{itemize}

Three distinct regions of $\delgs$ are apparent in Fig.~\ref{tevdiscovery}.
\begin{itemize}
\item
When $|\delgs|$ is large the upper limits on $\mgl$
for which the signal can be detected become independent of
$\delgs$ and asymptote to those for which $\gl\gl$ pair production could
be observed in the universal boundary condition scenario
for a given set of cuts (taking squarks to be much heavier than the $\gl$). 
This asymptote is reached already for $-\delgs\gsim 10$ since
for such values $\mgl/\mcpmone\sim 3$ (as for universal boundary
conditions), despite the fact that
$\mcpmone/\mcnone$ does not reach the universal
boundary condition result of $\sim 2$ even by $\delgs=-20$.
\item
When $|\delgs|$ is  small, the ratio $\mgl/\mcpmone$ exceeds the 
value $\sim 3$ typical of the universal boundary condition scenario,
and even higher values of $\mgl$ can be probed due to the increased energy
of the jets from $\gl$ decay, coupled with the fact that there is no loss
of $jets+\etmiss$ signal from the restriction against isolated leptons.
\item
For $\delgs$ such that $\mgl-\mcpmone$ and $\mgl-\mcnone$ are
small, the reach in $\mgl$ is reduced, although perhaps
less severely than naively anticipated.  The reason for this latter
is that even for small mass difference, it is still
possible to get energetic jets from initial and final state radiation rather
than from the $\gl$ decay. For most events at least
one of the three most energetic jets is radiative in
nature when $\mgl\sim\mchi$,  whereas for the universal boundary
condition scenario the most energetic jets are almost always
from the $\gl$ decays. 
\end{itemize}

\begin{table}[hbt]
\caption[fake]{Maximum $\mgl$ values that can be probed
using D0 and CDF cuts (see text) in the $jets+\etmiss$ final state
for different integrated luminosities, $L$, at the Tevatron and Tev*
at $\delgs=-10$, $\delgs=-1$ and $\delgs=-4.5$.
Observability is defined by $S/\sqrt B\geq 5$ and $S/B\geq 0.2$.
Also given are the maximum $\mgl$ values for which $\gl\gl$
production can be observed in the (universal boundary
condition) limit of very large $|\delgs|$
using the stronger cuts of Ref.~\cite{baeretal}. The results of this
table are for $\tanb=2$ and $\mu<0$.}
\begin{center}
\begin{tabular}{|c|l|rrrrr|}
\hline
Cuts & 
\ \ \ \ \ $L=$ & $8\pbi$ & $19\pbi$ & $100\pbi$ & $2\fbi$ & $25\fbi$ \\
\hline
\ & $\delgs=-1$ & $170\gev$ & $200\gev$ & $230\gev$ & $250\gev$ & $250\gev$ \\
D0 & $\delgs=-4.5$ & ---\ \ \ \ \  & $80\gev$ & $110\gev$ & $150\gev$ & $150\gev$ \\
\ & $\delgs=-10$ & ---\ \ \ \ \  & $140\gev$ & $200\gev$ & $200\gev$ & $200\gev$ \\
\hline
\ & $\delgs=-1$ & $180\gev$ & $190\gev$ & $230\gev$ & $250\gev$ & $250\gev$ \\
CDF & $\delgs=-4.5$ & $100\gev$ & $130\gev$ & $140\gev$ & $160\gev$ & $160\gev$ \\
\ & $\delgs=-10$ & $150\gev$ & $170\gev$ & $210\gev$ & $240\gev$ & $240\gev$ \\
\hline
Strong & $\delgs\to -\infty$ & \multicolumn{3} {c} {$\mgl$ excluded
by $\mcpmone\leq 47\gev$} & $250\gev$ & $300\gev$ \\
\hline
\end{tabular}
\end{center}
\label{maxmgl}
\end{table}

In Table~\ref{maxmgl} we give the maximum
$\mgl$ values that can be probed in the $jets+\etmiss$ channel,
using the D0 or CDF jet cuts delineated earlier, at $\delgs=-1$, 
$\delgs=-4.5$ (near the $\mgl\sim\mcpmone$ boundary)
and $\delgs=-10$. Notice
that there is no gain in discovery reach in going from $L=2\fbi$
(typical of the main injector) to Tev* luminosity of $L=25\fbi$.
This is due to the fact that the maximum $\mgl$ values
that can be probed at $L=2\fbi$ are determined by $S/B$ falling
below the minimum value of $0.2$.  For both the D0 and CDF 
cuts $B$ is relatively
large. If systematic uncertainties
in the predicted level of the $jets+\etmiss$ signal due
to theoretical and experimental uncertainties can be reduced
below the $\sim 10\%$ level, higher values of $\mgl$ could be probed.

The most striking difference between the D0 and CDF cuts is the 
much greater sensitivity of the softer CDF cuts to parameter
choices for which $\mgl\sim\mcpmone\sim\mcnone$.  
The reason for this striking difference 
is that the D0 cuts include a fairly stiff
minimum $E_T=25\gev$ requirement for the 3rd jet. 
As shown in Fig.~\ref{spectrumcompare}, for scenarios with
a small $\mgl$~--~$\mchi$ mass splitting this will eliminate
a substantial fraction of the signal events. Weakening this cut,
as in the CDF procedure, increases the signal rate, and apparently 
does so without increasing the background rate (perhaps
because stronger rapidity cuts are imposed on the jets in the CDF
procedure). It will be important
for the CDF and D0 collaborations to determine if there
are still more optimal cuts for scenarios with small mass splitting.
We note
that the greater sensitivity of CDF cuts persists up to $|\delgs|\sim 10$,
even though $\mgl-\mcpmone$ approaches values
typical of universal boundary conditions. 
In contrast, the difference between D0 and CDF cuts disappears
at small $|\delgs|$ where $\mgl-\mcpmone$ is even larger
than predicted for universal boundary conditions.

It is important to note that the jet cuts employed here are no longer optimal
in the universal boundary condition limit (roughly $-\delgs\gsim 10-20$)
when $L>2\fbi$ and $\mgl$ is large.  It is better to strengthen
the cuts. This is due to the fact that stronger
cuts will reduce the background rate $B$ while leaving
good efficiency for the signal if $\mgl$ is large and there
is substantial mass splitting between the $\gl$ and the $\cpmone$ and $\cnone$.
This leads to $S/B>0.2$ at high $\mgl$.
The stronger (more optimal) cuts employed at high $L$ are discussed
in Ref.~\cite{baeretal}.  For the $jets+\etmiss$ channel they are:
\begin{itemize}
\item  $\etmiss>40\gev$;
\item $n(jets)\geq 2$ jets having $|\eta_{\rm jet}|<3$ and $E_T(jet)>15\gev$;
\item Transverse sphericity $S_T\geq 0.2$;
\item $\Delta\phi(\etmiss,j(E_T>15\gev))>30^\circ$;
\item $E_T(j_1),E_T(j_2)\geq E_T^c$ and $\etmiss\geq E_T^c$,
with $E_T^c$ optimized for given $L$, $\mgl$ and other and parameter choices.
\end{itemize}
To give an example, the $\mgl=300\gev$ detection limit in Table~\ref{maxmgl}
for universal boundary conditions is attained using $E_T^c=100\gev$.
It is also possible that the stronger cuts would allow
us to probe to higher $\mgl$ values at small $|\delgs|$
than accessible using the weaker D0 and CDF jet cuts 
(see Fig.~\ref{tevdiscovery}). This is because
$\mgl/\mcpmone\sim \mgl/\mcnone$ is large and the jets would be energetic.
However, we have not performed a detailed analysis.

The limitations on the $\mgl$ discovery reach in the case of
$\mgl$ near $\mchi$ are not quite as much of a concern as
one might first suppose.  This is because the $m_0\lsim 2\tev$
naturalness requirement imposes an upper bound of $\mgl\lsim 140\gev$
in the $\mgl\sim \mcpmone$ boundary region, see Fig.~\ref{mglcontours}.
From Fig.~\ref{tevdiscovery} and Table~\ref{maxmgl} we see that this
value of $\mgl$ can be probed at $\delgs=-4.5$ with $L=100\pbi$,
provided CDF-like cuts are employed.
In contrast, for $\delgs=-10$, from Fig.~\ref{mglcontours} 
$\mgl\gsim 350\gev$ if $m_0=2\tev$ and $\gl\gl$ pair production
would not be observable even at Tev*.

Turning now to the LHC, the standard cuts employed there
in the universal boundary condition model are \cite{baeretalold}:
\begin{itemize}
\item There are no isolated leptons with $E_T>20\gev$, where
isolation is defined by requiring that additional $E_T$ within
$\Delta R\leq 0.3$ of the lepton be $<5\gev$. 
\item Transverse sphericity of $S_T>0.2$.
\item $\etmiss>500\gev$.
\item $n(jets)\geq 2$ having $|\eta_{\rm jet}|<3$
and $E_T>100\gev$ using coalescence within $\Delta R\leq0.7$.
These are ordered according to decreasing $p_T$.
\item $30^\circ\leq \Delta\phi(\etmiss,jet)<90^\circ$ for the
jet which is closest to the $\etmiss$ vector.
\item $E_T(j_1,j_2)\geq 500\gev$.
\end{itemize}
For these cuts, $\gl\gl$ pair production in the O-II model is not
observable in any part of the $|\delgs|<10$
portion of parameter space appearing in Fig.~\ref{tevdiscovery}.
Hopefully, this will be cured by weakening the jet cuts.
Optimization of the cuts and assessment of the signals is underway
and will be presented elsewhere.

\subsubsection{Leptonic Signals for {\boldmath $\gl\gl$} Production}

We have already remarked that if $\mcpmone\sim\mcnone$, then
it will be far more difficult to extract a
like-sign di-lepton signal for $\gl\gl$ production than
is the case for universal boundary conditions.
The leptons from the $\cpmone$ decays will
be much softer when $\dmchi$ is small. A detailed study
will be required to determine if any signal survives.
We believe it is unlikely that the di-lepton signal
can achieve as much discovery reach as the $jets+\etmiss$ signal.

When $\mgl$ is close to $\mchi$, there are
no other sources of leptons than those from the $\cpmone$'s
that are present in the $\gl$ decays. As $\mgl$ becomes larger
than $\mchi$, not only does the $jets+\etmiss$
signal become increasingly strong, but also additional
leptonic signals for $\gl\gl$ production emerge deriving
from $\gl\to jets+\cntwo$ decay followed by $\cntwo\to \ell^+\ell^- \cnone$,
dominated by the on-pole $Z\cnone$ final state when kinematically allowed.
A detailed study is required to determine if the resulting
signal for $\gl\gl$ production is competitive with the $jets+\etmiss$
signal.  Ultimately, in the very large $|\delgs|$ universal boundary 
condition limit, the presence of the many cascade decays by which $\gl\gl$
production leads to leptons results in a $\ell^\pm+jets+\etmiss$
signal that is stronger than the $jets+\etmiss$ signal at the LHC
\cite{baeretalold}.

\subsubsection{The Tri-Lepton Signal}

We have also performed explicit simulations for the tri-lepton signal
at the Tevatron. A summary of current CDF and D0 cuts
and results appears in Ref.~\cite{tkamon}. In our analysis
we consider two sets of cuts.
The first set of cuts is that employed by CDF in analyzing
their $L=100\pbi$ data set:
\begin{itemize}
\item $|\eta(\ell_{1,2,3})|\leq 2.5$;
\item $E_T(\ell_1)>11\gev$, $E_T(\ell_{2,3})>4\gev$;
\item $\etmiss>15\gev$;
\item $n(jets)=0$ for jets with $E_T>15\gev$;
\item events with $\epem$ or $\mupmum$ pairs with mass $\sim\mz$
are vetoed.
\end{itemize}
The second set of cuts \cite{baeretal} was designed to
detect $3\ell$ events at luminosities $\gsim 1\fbi$
(Main Injector and Tev*) in the universal boundary condition scenario.
The latter cuts (which we call the `strong' cuts) are:
\begin{itemize}
\item $|\eta(\ell_{1,2,3})|\leq 2.5$;
\item $E_T(\ell_{1,2,3})\geq 20,15,10\gev$, respectively;
\item $\etmiss\geq 25\gev$;
\item $n(jets)=0$ for jets with $E_T>15\gev$;
\item events with $\epem$ or $\mupmum$ pairs with mass $\sim\mz$
are vetoed.
\end{itemize}
For the CDF cuts the background was taken from Ref.~\cite{tkamon}
as 0.4 events for $L=100\pbi$, corresponding to a cross section
of $\sigma_B=4\fb$. For other luminosities the number of
background events was computed as $L\sigma_B$.
For the second set of cuts, the background was explicitly computed using
ISAJET, summing over all important reactions.
For both sets of cuts,
the $3\ell$ signal is deemed observable if there are at least 5
events, $S/\sqrt B\geq 5$ and $S/B\geq 0.2$.

For either set of cuts,
the $3\ell$ signature is unobservable for any luminosity
unless $\delgs$ is well above the region where $\dmchi$ is small.
At $\delgs=-9$, the $3\ell$ remains unobservable for both
sets of cuts at $L=100\fbi$. 
For the CDF (strong) cuts, a $3\ell$ signal becomes
observable at $\delgs=-9$ for $\mgl$ values up to $160\gev$
($100\gev$) with $L=2\fbi$ and up $210\gev$ ($160\gev$) 
with $L=25\fbi$.  The fact that the strong cuts do not allow
as much sensitivity to the $3\ell$ signal as do the weaker CDF cuts
is obviously a result of the fact that the leptons remain soft (as compared
to universal boundary condition expectations)
in the O-II model out to quite large $|\delgs|$.

\subsubsection{Gluino-Squark and Squark-Squark Pair Production}

Production of $\gl\sq$ and $\sq\sq$ pairs will occur at a significant
rate at the LHC, even for squark masses of $1\tev$ or more.
These pair production processes would be followed by
$\sq\to \gl q$ decay, leading to final states comprised of $\gl\gl q$
and $\gl\gl q q$, respectively.  The $\gl$'s would then decay
as we have described. In particular, it is very possible that
most of the energy of the $\gl$ will go into the $\cnone$.
Thus, even though the $\gl$'s from the $\sq$ decays would be
energetic, the visible jet energy component of the decay need
not be large. Further, it will tend to be aligned with the missing
energy component due to the large momentum of the $\gl$
coming from decay of a very heavy $\sq$.

Thus, $\gl\sq$ production will lead to a final state with
a very energetic quark and large missing energy 
in the opposite direction. Backgrounds to this configuration need
to be studied to determine if the signal
for such events can be found. (An obvious background is $Z+g$
production in which the $Z$ decays invisibly.)
We have not attempted this study here.
The $\sq\sq\to q q \gl\gl\to q q \etmiss+soft$ signal for SUSY
is not very different than that already considered 
(see, \eg, Ref.~\cite{baeretalold}) for $\sq\sq$
production in the case of universal boundary conditions.
The final state would consist of
of two highly energetic jets along with large
$\etmiss$, all in different directions; backgrounds will be small,
and detection of the signal should be straightforward for $\msq\lsim 1.5$ to
$2\gev$ at the LHC \cite{baeretalold,ATLAS,CMS,womersley}.
For both $\gl\sq$ and $\sq\sq$ production, it could be that
stop and sbottom squarks would be easiest to trigger on
due to the fact that the
final states would contain two $t$'s or two $b$'s, respectively.
(Note that in $\gl\sq$ production with $\sq=\wtil t$ or $\wtil b$,
there must be an associated $\anti t$ or $\anti b$, respectively.)
In the case of $\gl\sq$ ($\sq\sq$) production, with $\sq=\wtil t$
or $\wtil b$, one (both) 
of the $t$'s or $b$'s would be very energetic.
Vertex $b$-tagging could be used to isolate the relevant events.
A close examination of these signals is warranted.

\subsection{Implications for Cosmology}

One of the attractive features of supersymmetry is that the
LSP could provide a natural source for the dark matter that appears
to be required by galactic rotational data (requiring
$\Omega h^2>0.025$) and that would be needed for closure of
the universe ($\Omega=1$); further, $\Omega h^2\lsim 1$ is
required in order that the universe be at least 10 billion years old.
(Here, $\Omega$ is the present LSP mass density in units of
the critical or closure density, and $\h$ is the Hubble constant in units
$100~{\rm km}/({\rm sec}\cdot{\rm Mpc})$.)
The $\mcpone\simeq\mcnone$ degeneracy and SU(2)-gaugino nature
of the $\cnone$ predicted
in the $\sin\theta\to 0$ O-II model when $|\delgs|$ is not
large leads to a picture
that differs substantially from that found for universal boundary conditions.
The two key differences are easily summarized:
\begin{itemize}
\item For moderate to small $|\delgs|$ in the O-II model,
$\cnone\cnone$ annihilation is quite small because
the $\cnone$ is usually almost pure SU(2) gaugino and its couplings 
to the light Higgs ($\hl$) and the $Z$ are weak (they require a higgsino
component) and because the sfermions are typically
very heavy [as discussed with regard to Eqs.~(\ref{smalld}) and (\ref{larged}),
and following]. 
\item
The near degeneracy $\mcpmone\simeq\mcnone$ implies
(as noted several years ago in Ref.~\cite{mngy})
similar densities (due to very similar Boltzmann factors)
for the $\cpmone$ and $\cnone$
at the time of freeze-out, so that co-annihilation between the LSP
and the chargino becomes very important and can greatly reduce the expected
relic density. 
\end{itemize}

\begin{figure}[htb]
\begin{center}
\centerline{\psfig{file=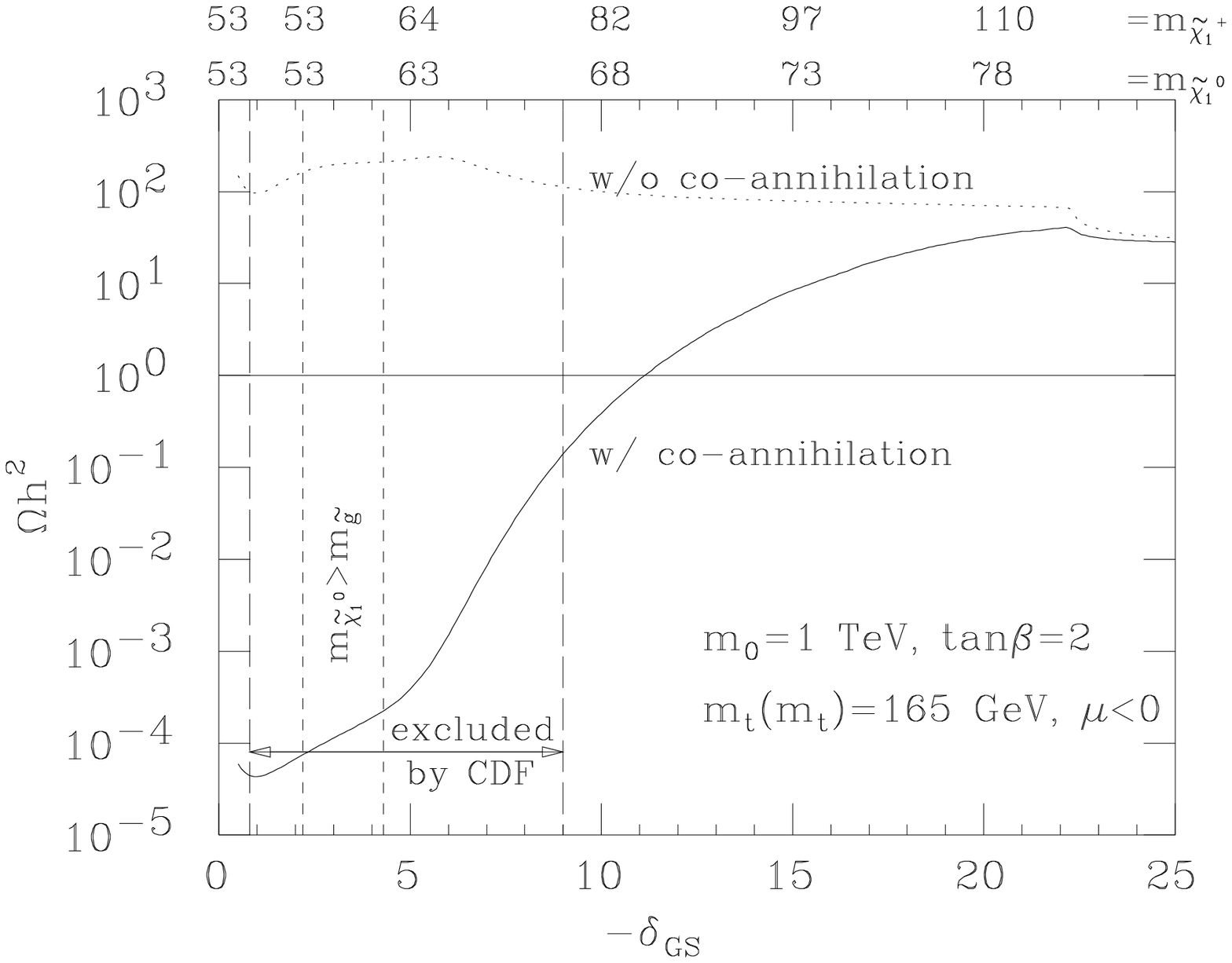,width=12cm}}
\bigskip
\begin{minipage}{12.5cm}       
\caption{We plot the relic density $\Omega h^2$ as a function
of $\delgs$ with and without including $\cpmone\cnone$ co-annihilation.
We take $m_0=1\tev$, $\tanb=2$, $\mt(\mt)=165\gev$ and $\mu<0$.
The region defined by the vertical short-dashed lines is disallowed
because the $\gl$ would be the LSP. The region defined by the 
vertical long-dashed lines is excluded by
the failure of the CDF collaboration to detect a $jets+\etmiss$
signal from $\gl\gl$ production for $L=19\pbi$ of data.
}
\label{coannihilation}
\end{minipage}
\end{center}
\end{figure}

The computation of the relic density is sketched in Appendix B.
As well as the $\cpmone\cnone\to f\anti f$ co-annihilation channel
considered in \cite{mngy}, we
also included the $\cpmone\cnone\to \wpm \gam$ channel (a few per cent
effect).
To illustrate the importance of co-annihilation, we have plotted
$\Omega h^2$, before and after including co-annihilation,
as a function of $\delgs$ (for $m_0=1\tev$, $\tanb=2$
and $\mu<0$) in Fig.~\ref{coannihilation}. From this figure,
we observe that without co-annihilation $\Omega h^2$ is at least 10,
and, at smaller $|\delgs|$, as large as 100, \ie\ drastically
inconsistent with the fact that the universe is still expanding.
After including co-annihilation, we see that $\Omega h^2<1$ for $|\delgs|\lsim
10$. Indeed, extremely small $\Omega h^2$ values, $\lsim 10^{-4}$, are
possible for $|\delgs|\lsim 6$ where $\dmchi$ becomes small.
As $|\delgs|$ increases above this, 
co-annihilation becomes ineffective when $\dmchi/\mcnone$ exceeds
a few \%. 

For the choice of parameters of Fig.~\ref{coannihilation},
$\Omega h^2<1$ requires $\mcpmone\lsim 88\gev$; direct detection
of $\cpone\cmone$ production at LEP-II will be possible
only for the portion of this range for which $\dmchi$ exceeds $5-10\gev$,
\ie\ roughly for $|\delgs|\gsim 7$, see Fig.~\ref{inosdmchi},
corresponding to $\mcpmone\gsim 70\gev$. (This is precisely
the range for which $\Omega h^2\gsim 0.025$, and the $\cnone$
of the model could be the dark matter of the universe.)
For $\mcpmone\lsim 70\gev$,
it will become necessary to employ the $\gam\cpone\cmone$ mode,
see Fig.~\ref{eetogaminv}.  The region of  parameter space
currently eliminated by the failure to observe $\gl\gl$ production
at the Tevatron employing the published CDF analysis of
$L\sim 19\pbi$ of data is obtained by correlating $\mgl$ and $\delgs$
locations in Fig.~\ref{coannihilation} with the CDF
excluded regions in Fig.~\ref{tevdiscovery};
it is roughly $0.8\lsim -\delgs\lsim 9$, as indicated in
Fig.~\ref{coannihilation}.
(Because of stronger jet cuts and the
smaller amount, $L=8\pbi$, of analyzed data,
the currently published D0 analysis only excludes
$1\lsim -\delgs \lsim 1.6$.)
For all but the $|\delgs|<0.6$
portion of the $-\delgs\lsim 12$ range,
$\mgl$ is such that $\gl\gl$ production at the Tevatron could
be detected using the CDF cuts and analysis procedures outlined in
the preceding section applied to $L=100\pbi$ of 
data, see Fig.~\ref{tevdiscovery}.
Thus, for the choices $\tanb=2$ and $m_0=1\tev$, it is 
only for very small $|\delgs|$ that the model can be consistent
with an expanding universe if no signal for $\gl\gl$
production is found after analyzing Run-Ia+Run-Ib Tevatron data
using the CDF procedures employed for Fig.~\ref{tevdiscovery}.
However, to repeat, 
small values of $|\delgs|$ will not yield $\Omega h^2\gsim 0.025$.

\begin{figure}[htb]
\begin{center}
\centerline{\psfig{file=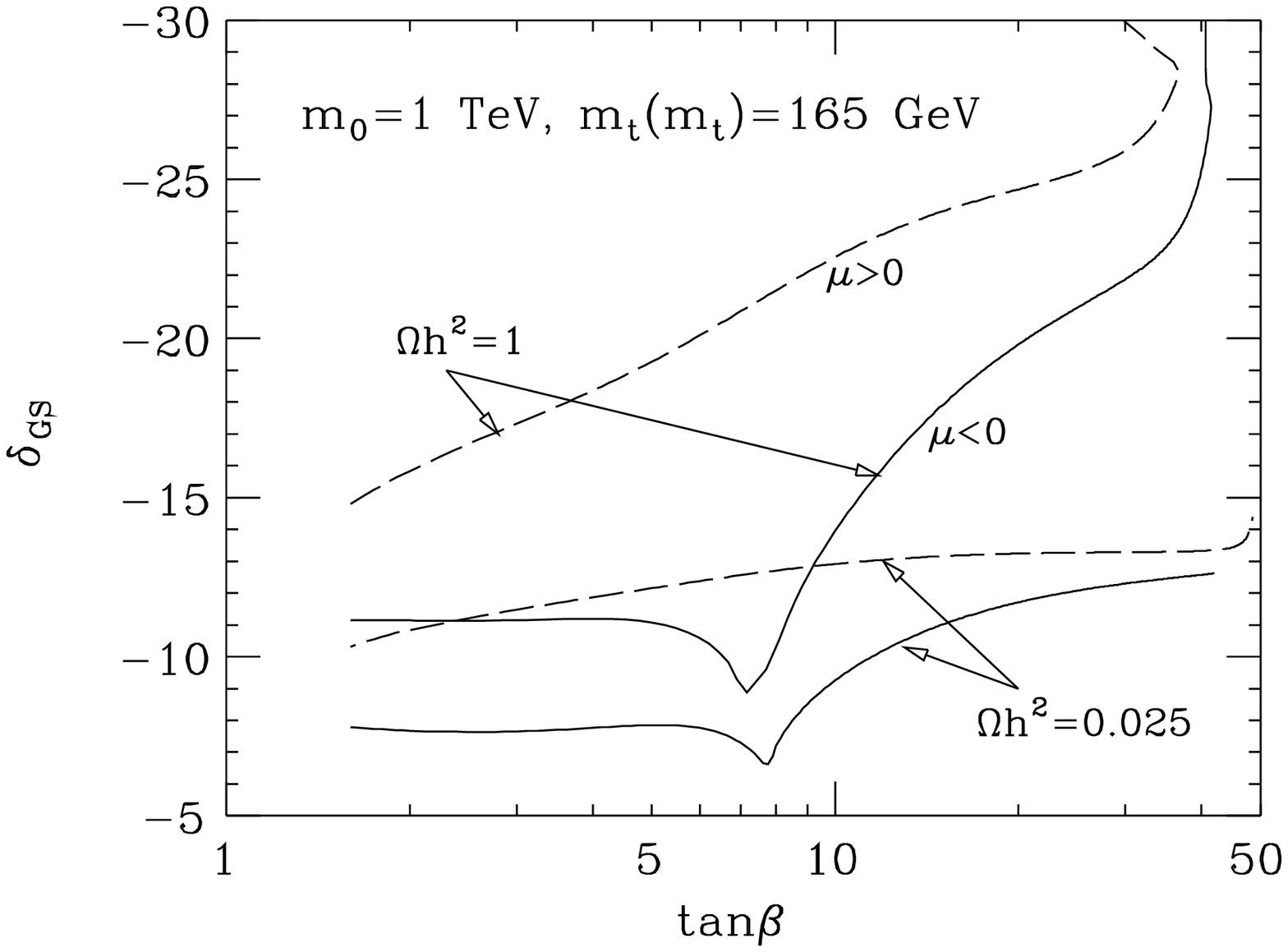,width=12cm}}
\bigskip
\begin{minipage}{12.5cm}       
\caption{Contours of constant $\Omega h^2=0.025$ and $\Omega h^2=1$
in $\tanb$--$\delgs$ parameter space at $m_0=1\tev$, for
$\mu<0$ and $\mu>0$.
}
\label{omegahsqdgstanbcontours}
\end{minipage}
\end{center}
\end{figure}

The range of $\delgs$ for which $0.025\lsim\Omega h^2\lsim 1$, especially
the largest allowed $|\delgs|$ value, typically increases
with increasing $\tanb$.
The regions with $0.025\leq \Omega h^2\leq 1$
are plotted in the $\delgs$--$\tanb$ parameter space plane
for $m_0=1\tev$ and both signs of $\mu$ in Fig.~\ref{omegahsqdgstanbcontours}.
Typically, only a narrow range of $\delgs$ values satisfies
both criteria unless $\tanb$ is very large.
The lower bound on $-\delgs$, set by requiring $\Omega h^2\geq 0.025$
would not be present if some other explanation for dark matter
is assumed to exist.  Indeed, for the preferred model
values of $\delgs=-4,-5$ an alternative source of dark matter would 
be necessary.

\begin{figure}[htb]
\begin{center}
\centerline{\psfig{file=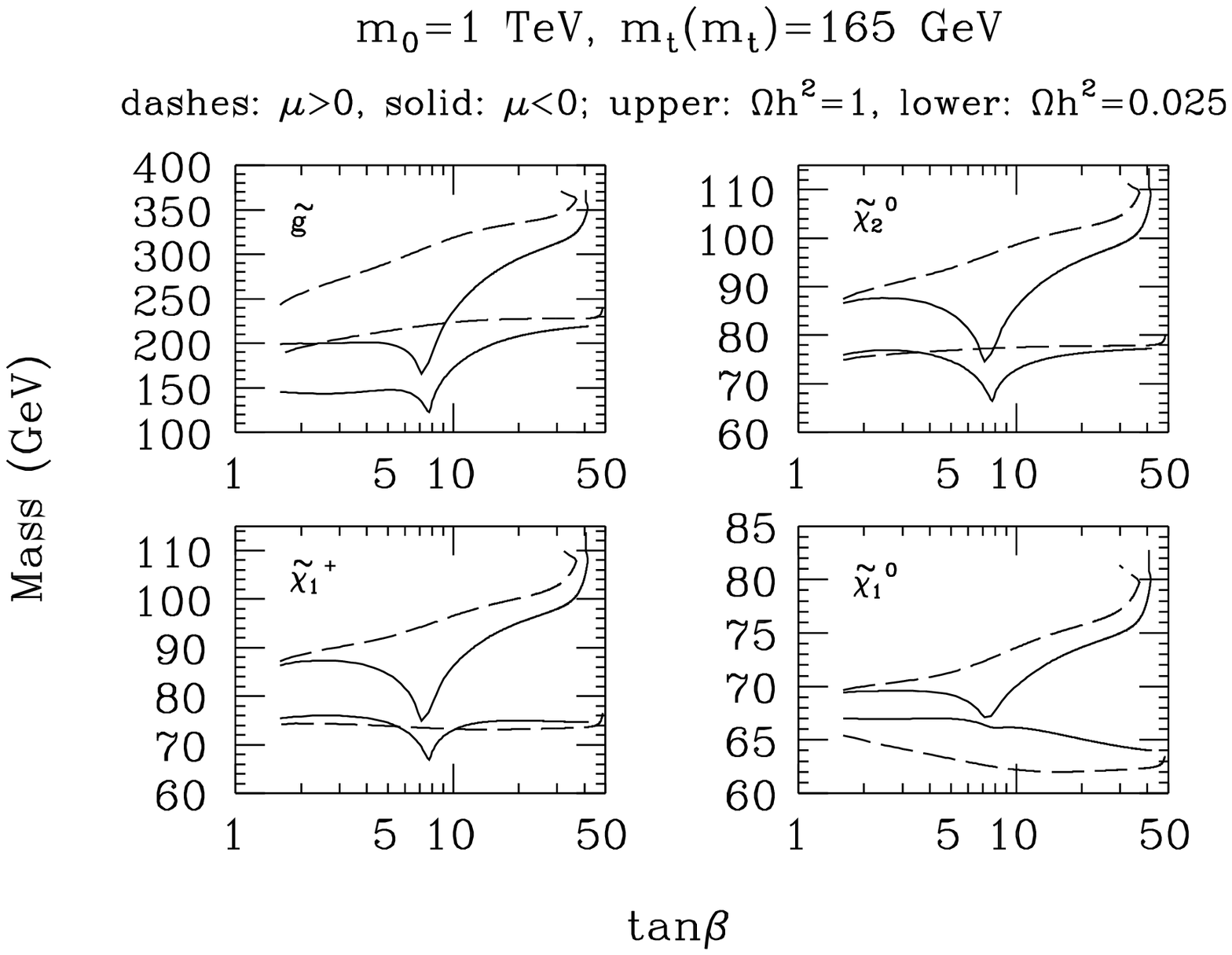,width=12cm}}
\bigskip
\begin{minipage}{12.5cm}       
\caption{Contours of constant $\Omega h^2=0.025$ and $\Omega h^2=1$
in the $\tanb$--$\mgl,\mcpmone,\mcnone,\mcntwo$ parameter spaces 
at $m_0=1\tev$, for $\mu<0$ and $\mu>0$.
}
\label{omegahsqmasstanbcontours}
\end{minipage}
\end{center}
\end{figure}

To further delineate the consistency of existing and near future experimental 
data with the constraints on $\Omega h^2$, it is illuminating to
plot the $\Omega h^2=0.025$ and $1$ contours in the $\mgl$--$\tanb$,
$\mcpmone$--$\tanb$, $\mcnone$--$\tanb$ and $\mcntwo$--$\tanb$
parameter spaces. That is, we simply convert from $\delgs$ to one of the
indicated masses.  These four sets of contours appear in
Fig.~\ref{omegahsqmasstanbcontours}. We observe that,
for small to moderate $\tanb$ values, the upper
bounds on the masses set by $\Omega h^2<1$ are such that the $\cpmone$
and $\cntwo$ should be observable at LEP-II 
(keeping in mind that the maximum masses occur for large $-\delgs$ for
which $\dmchi$ is big enough that direct $\cpone\cmone$
detection should be feasible) and
such that $\gl\gl$ detection at the Tevatron should be possible.
At larger $\tanb$ ($\gsim 5-10$, depending upon $\sign(\mu)$),
the masses begin to exceed the reach of LEP-II and the Tevatron.
Large masses for the gauginos are also possible if $-\delgs$ is very small.
However, as noted already, if we require $\Omega h^2>0.025$ then
$-\delgs$ can never be small (see Fig.~\ref{omegahsqdgstanbcontours}). 
Together, the $\Omega h^2>0.025$ and $\Omega h^2<1$ bounds imply
that the gaugino masses must all lie in mass regions that
are eminently accessible at the NLC and LHC, and very possibly at LEP-II 
and the Tevatron.\footnote{This is not dissimilar to the conclusion that
is reached in the case of universal boundary conditions if
$m_0$ is large. There, if $m_0$ is large enough ($\gsim 300\gev$)
to suppress $t$-channel annihilation contributions
to $\cnone\cnone$ annihilation, then the $\cnone$ must be light
enough that annihilation via a nearby $s$-channel Higgs and/or $Z$ pole
is sufficiently efficient. Typically \cite{lopezetal}, $\mcnone\lsim 55\gev$
or so is required.}
With regard to the higgsino-like chargino and neutralinos, we
refer back to Fig.~\ref{sigcontours}, where
the $\Omega h^2=0.025$ and $1$ contours were given.
We observe that
if $\tanb\gsim 4-5$ then higgsino discovery at a $\rts=500\gev$
$\epem$ collider will generally be possible for model parameters
consistent with $0.025\lsim \Omega h^2 \lsim 1$. 

Finally, we note
that the $\Omega h^2$ contours are not sensitive
to the $m_0$ value if $m_0\gsim 200\gev$ so that the co-annihilation
cross section is mainly determined by the $s$-channel $W$ pole graph.
Thus, even though $m_0$ could be substantially below
$1\tev$ for small $-\delgs$ 
without violating $\mcpmone\gsim 47\gev$ (due to the $m_0/\sqrt{-\delgs}$
growth of the $|M_i^0|$'s), small $-\delgs$ values would continue
to be ruled out if $\Omega h^2\gsim 0.025$ is required.

\section{Final Remarks and Conclusions}

The moduli dominated limit of string SUSY breaking yields a rich
phenomenology that differs substantially from that obtained
for the usual $\mgut$-scale universal boundary conditions.
In this paper we have considered
a specific orbifold model (the O-II model)
in which the moduli dominated limit can be taken and the $\mgut$-scale
boundary conditions computed.
The model and its phenomenology are determined by $\tanb$ (the
standard Higgs vacuum expectation value ratio), $\delgs$
(the Green-Schwarz mixing parameter) and the universal scalar mass at $\mgut$,
$m_0$.  Theoretically, negative integer values for $\delgs$
in the range $|\delgs|<5-6$ are preferred.  For such values,
the gaugino masses at $\mgut$, which only arise at one-loop, 
are very non-universal; universality is
approached, but {\it very} slowly, as $|\delgs|$ becomes very large.
Further, the (one-loop) $\mgut$-scale gaugino masses, $|M_i^0|$,
 will be very much smaller
than $m_0$. The non-universality of the gaugino masses at $\mgut$ implies that
it is very possible that the lightest chargino and neutralino
will both be SU(2) gauginos and, therefore,
approximately degenerate, $\mcpmone\simeq\mcnone$.
Further, it is also possible for the gluino to be degenerate with both
just outside the region $\delgs\sim -3$ that is excluded by
virtue of requiring $\mgl\geq \mcnone$, \ie\ that the gluino
not be the LSP. If $|\delgs|$ is of moderate size, implying $m_0\gg|M_i^0|$, 
the gauginos will be relatively light provided $m_0\lsim 2\tev$ (as
presumably required by naturalness); indeed, 
$m_0\gsim 1\tev$ if $\mcpmone\gsim \mz/2$,
as required by LEP data. When $m_0$ is large, squarks, sleptons and
heavy Higgs bosons will be very massive. Further, correct
electroweak symmetry breaking implies that $\mu$ will be large.
Thus, the most
accessible SUSY signals will be those deriving from gaugino ($\gl$,
$\cpmone$, $\cnone$ and $\cntwo$) production.

The mass degeneracies noted above are of particular
phenomenological importance.
Key implications at existing and future accelerators include
the following.
\begin{itemize}
\item If $\mcpmone-\mcnone$ 
is small (as for theoretically preferred
model parameters) it will be necessary 
to employ $\epem\to\gam\cpone\cmone$ final
states at LEP-II and the NLC for light chargino detection
due to the near invisibility of the $\cpmone$ decays.
\item The small size of $\epem\to \cnone\cntwo,\cntwo\cntwo$
cross sections (due to the state compositions predicted
by the model) and the large masses of the $\cpmtwo$, $\cnthree$
and $\cnfour$ (due to the large value of $|\mu|$ predicted),
imply that only at the NLC can one hope for substantial
numbers of neutralino and chargino pair events other than
the difficult to detect $\cpone\cmone$ process.
\item If $\mgl\sim \mcpmone\simeq\mcnone$ (as for theoretically
preferred model parameters), $\gl\gl$ production will be more difficult
to detect at both the Tevatron and the LHC due to the softness
of the jets in $\gl$ decay. Weak jet cuts must
be employed, implying large background and difficulty
in achieving adequate $S/B$.
\item
If $\mgl$ is not nearly degenerate with $\mcpmone\simeq\mcnone$,
$\gl\gl$ discovery in the $jets+\etmiss$ channel 
(using weak jets cuts) at the Tevatron/Tev* 
could be easier than observation of neutralino and chargino
pair production at a $\rts=500\gev$ NLC.  
\item Detection of $\gl\gl$ production at the LHC will require
significant alterations in the cuts currently employed.
\item If $\mcpmone\simeq\mcnone$, the degeneracy will be manifest at 
a hadron collider as an absence of like-sign
dilepton signals for $\gl\gl$ production and of tri-lepton signals
for $\cpmone\cntwo$ production. 
\end{itemize}
Cosmological constraints on the model are significant.
\begin{itemize}
\item
For the preferred $|\delgs|\lsim 5-6$ range, $\Omega h^2<0.025$
(the minimum required if the $\cnone$ is to be a significant dark
matter candidate). Such small values of $\Omega h^2$ are a
result of the large $\cpmone\cnone$ co-annihilation rate when
$\mcpmone\simeq\mcnone$. 
\item
If $\delgs$ is chosen as a function of $\tanb$ so that $0.025\lsim \Omega
h^2\lsim 1$, then for moderate $\tanb$ the
$\gl$, $\cpmone$, and $\cntwo$ masses are relatively modest
in size.  In particular, they are such that
proper analysis of existing Tevatron data 
and soon-to-come LEP-II data will exclude $\tanb\lsim 5-10$.
\end{itemize}

\section{Acknowledgements}

This work was supported in part by U.S. Department of Energy grants
DE-FG03-91ER40674 (JFG, CHC) and DE-FG02-95ER40896 (MD), 
the Davis Institute for High Energy Physics, the Wisconsin Research Committee
using funds granted by the Wisconsin Alumni Research Foundation (MD),
and by a grant from the Deutsche Forschungsgemeinschaft under
the Heisenberg program (MD).

\section*{Appendix A: Chargino Decays}
\renewcommand{\theequation}{A.\arabic{equation}} 
\setcounter{equation}{0}
In this Appendix we discuss the calculation of the partial widths for leptonic
and hadronic chargino decays for small mass splitting between the chargino and
the lightest neutralino. We saw in Fig.~\ref{inosdmchi} 
that even after inclusion of
1--loop radiative corrections \cite{mngy,pappierce} 
this mass difference can be as
small as 150 MeV. Standard expressions for $\tilde{\chi}_1^\pm \rightarrow
\tilde{\chi}_1^0 f \bar f'$ are not applicable for such small mass differences,
since they assume the final state fermions $f$ and $\bar f'$ to be massless.
Moreover, hadronic decays can only be described by perturbative QCD if the
mass difference exceeds one or two GeV. 

Allowing a finite mass for the standard model fermions produced in chargino
decays is straightforward. First, we can ignore sfermion and charged Higgs
exchange diagrams, since these particles are very heavy, and the 
$\tilde{\chi}_1^\pm \tilde{\chi}_1^0 W$ coupling is maximal in the relevant
limit where both the chargino and the neutralino are almost pure $SU(2)$
gauginos. Further, we actually only need to keep the mass of one of the two 
SM fermions; the other one is either much lighter (as for $f=c,\ f'=s$)
or exactly massless (for $f=l, \ f' = \nu_l$). The result can be written as:
\begin{eqnarray} \label{ea1}
\Gamma(\tilde{\chi}_1^- \rightarrow \tilde{\chi}_1^0 & & f \bar{f}') =
\frac {N_c G_F^2} {(2 \pi)^3} 
\Biggl\{ \widetilde{m}_- \left[ \left( O_{11}^L \right)^2
+ \left( O_{11}^R\right)^2 \right] 
\nonumber \\
\cdot & &
~\int_{(\widetilde{m}_0+m_f)^2}^{\widetilde{m}_-^2}
d q^2 \left( 1 - \frac {\widetilde{m}_0^2 + m_f^2} {q^2} \right)
\left( 1 - \frac {q^2} {\widetilde{m}_-^2} \right)^2 
\sqrt{\lambda(q^2,\widetilde{m}_0^2,m^2_f)} 
\\
 & & 
\hspace*{-5mm} - 2 \widetilde{m}_0 O^L_{11} O^R_{11} 
\int_{m^2_f}^{(\widetilde{m}_--\widetilde{m}_0)^2} dq^2 \frac {q^2} 
{\widetilde{m}^2_-}
\left( 1 - \frac {m_f^2}{q^2} \right)^2 
\sqrt{\lambda(\widetilde{m}^2_-,\widetilde{m}^2_0,q^2)} \Biggr\}\, . 
\nonumber 
\end{eqnarray}
Here, $\lambda(a,b,c) = (a+b-c)^2-4ab$ is the standard kinematical function,
$N_c = 3 \ (1)$ if $f$ is a quark (lepton), 
$O^{R,L}_{11}$ are the $\tilde{\chi}_1^\pm \tilde{\chi}_1^0 W$ couplings
in the notation of Ref.~\cite{kanehaber}, and we have introduced the
shorthand notation $\widetilde{m}_0 \equiv m_{\tilde{\chi}_1^0}, \ 
\widetilde{m}_- \equiv m_{\tilde{\chi}_1^\pm}$. In the limit where both
$\tilde{\chi}_1^\pm$ and $\tilde{\chi}_1^0$ are pure $SU(2)$ gauginos,
$O^L_{11} = O^R_{11} = 1$.

As mentioned earlier, Eq.~(\ref{ea1}) can only describe hadronic
$\tilde{\chi}_1^\pm$ decays if the chargino--neutralino mass
difference $\Delta m_{\tilde \chi}$ is sufficiently large. For $\Delta
m_{\tilde \chi} < 1$ to 2 GeV one instead has to explicitly sum over
exclusive hadronic final states. Fortunately much work on the related
case of semi--leptonic $\tau$ decays has already been done. We
adopt the formalism developed in Ref.~\cite{kuhn}.

As stated in the main text, we must include the partial widths for
$\tilde{\chi}_1^- \rightarrow \pi^- \tilde{\chi}_1^0$,
$\cmone\to \pi^-\pi^0\cnone$, $\cmone\to \pi^-\pi^0\pi^0\cnone$
and $\cmone\to \pi^0\pi^+\pi^0\cnone$. As already noted,
the single pion mode will dominate for $m_\pi<\dmchi\lsim 1\gev$.
Partial widths into final states containing kaons are suppressed by a factor
of $\sin^2 \theta_c \simeq 1/20$. 
The relevant partial widths can be written as \cite{kuhn}
\ben \label{en2} \beq
\Gamma(\tilde{\chi}_1^- \rightarrow \tilde{\chi}_1^0 \pi^-)
&= \frac {f_\pi^2 G_F^2} {4 \pi} \frac {|\vec{k}_\pi|}{\widetilde{m}_-^2}
\left\{ \left( O^L_{11} + O^R_{11} \right)^2 \left[ \left(
\widetilde{m}_-^2 - \widetilde{m}_0^2 \right)^2 - m^2_\pi \left(
\widetilde{m}_- - \widetilde{m}_0 \right)^2 \right]
\right. \nonumber \\ & \left.
\hspace*{25mm} + \left( O^L_{11} - O^R_{11} \right)^2 \left[ \left(
\widetilde{m}_-^2 - \widetilde{m}_0^2 \right)^2 - m^2_\pi \left(
\widetilde{m}_- + \widetilde{m}_0 \right)^2 \right] \right\}; \\
\label{en2a}
& \phantom{nothing}\nonumber \\
\Gamma(\tilde{\chi}_1^- \rightarrow \tilde{\chi}_1^0 \pi^- \pi^0)
&= \frac {G_F^2} {192 \pi^3 \widetilde{m}_-^3} 
\int_{4 m_\pi^2}^{(\Delta m_{\tilde{\chi}_1})^2} d q^2 \left| F(q^2) \right|^2
\left( 1 - \frac {4 m_\pi^2}{q^2} \right)^{3/2}
\lambda^{1/2}(\widetilde{m}_-^2,\widetilde{m}_0^2,q^2)
\nonumber \\ & \hspace*{23mm}
\times \left\{ \left[ \left( O^L_{11} \right)^2 + \left( O^R_{11} \right)^2 \right]
\left[ q^2 \left( \widetilde{m}_-^2 + \widetilde{m}_0^2 - 2 q^2 \right)
+ \left( \widetilde{m}_-^2 - \widetilde{m}_0^2 \right)^2 \right]
\right. \nonumber \\ & \left. \hspace*{25mm}
- 12 O^L_{11} O^R_{11} q^2 \widetilde{m}_- \widetilde{m}_0 \right\}; \\
\label{en2b}
& \phantom{nothing}\nonumber \\
& \phantom{nothing}\nonumber \\
\Gamma(\tilde{\chi}_1^- \rightarrow \tilde{\chi}_1^0 3\pi)
&= \frac {G_F^2} {6912 \pi^5 \widetilde{m}_-^3 f_\pi^2} 
\int_{9 m_\pi^2}^{(\Delta m_{\tilde{\chi}_1})^2} d q^2 
\lambda^{1/2}(\widetilde{m}_-^2,\widetilde{m}_0^2,q^2)
\left| BW_a(q^2) \right|^2 g(q^2)
\nonumber \\ & \hspace*{26mm}
\times\Bigg\{ \left[ \left( O^L_{11} \right)^2 + \left( O^R_{11} \right)^2 \right]
\left[ \widetilde{m}_-^2 + \widetilde{m}_0^2 - 2 q^2
+ \frac {\left( \widetilde{m}_-^2 - \widetilde{m}_0^2 \right)^2} {q^2} \right]
\nonumber \\ & \hspace*{28mm}
- 12 O^L_{11} O^R_{11} \widetilde{m}_- \widetilde{m}_0 \Bigg\}.
\label{en2c}
\eeq \een 
In Eq.~(\ref{en2}a), $|\vec{k}_\pi|
= \lambda^{1/2}(\widetilde{m}_-^2, \widetilde{m}_0^2, m^2_\pi)/(2
\widetilde{m}_-)$ is the pion's 3--momentum in the
chargino rest frame, and $f_\pi \simeq 93$ MeV is the pion decay
constant. The form factor $F(q^2)$ 
in Eq.~(\ref{en2}b) is dominated by the $\rho$ and $\rho'$ meson poles:
\begin{equation} \label{ea3}
F(q^2) = \frac{ BW_{\rho}(q^2) + \beta BW_{\rho'}(q^2) } {1 + \beta}\,.
\end{equation}
Here $BW$ stands for a Breit--Wigner pole:
\begin{equation} \label{ea4}
BW_V(q^2) = \frac {m_V^2} { m_V^2 - q^2 - i \sqrt{q^2} \Gamma_V},
\end{equation}
with $V = \rho, \ \rho'$. Following Ref.~\cite{kuhn} we use $\beta = -0.145$
in Eq.~(\ref{ea3}), and $m_{\rho} = 773$ MeV, $\Gamma_{\rho} = 145$ MeV,
$m_{\rho'} = 1370$ MeV and $\Gamma_{\rho'} = 510$ MeV in Eq.~(\ref{ea4}).
Explicit expressions
for the Breit--Wigner propagator $BW_a$ of the $a_2$ meson, the exchange of
which is assumed to dominate $3\pi$ production, as well as for the
three-pion phase space factor
$g(q^2)$ can be found in Eqs.~(3.16)-(3.18) of Ref.~\cite{kuhn}.  
We have used the
propagator without ``dispersive correction''. This underestimates the
partial width for $\tau^- \rightarrow 3 \pi \nu_\tau$ decays by about
35\%; in our numerical results, we have therefore multiplied the
r.h.s. of Eq.~(\ref{en2c}) by 1.35. Nevertheless the branching ratio
for these modes never exceeds $\sim 18\%$; note that Eq.~(\ref{en2c})
includes both $\pi^- \pi^0 \pi^0$ and $\pi^- \pi^- \pi^+$ modes, which
occur with equal frequency.                       

We use Eqs.~(\ref{en2}) 
to describe hadronic $\tilde{\chi}_1^\pm$ decays as long
the sum of the three partial widths is larger than predicted by
Eq.~(\ref{ea1});  in the latter, we
use a constituent-type effective mass for the $d-$quark of 500 MeV in our 
calculation (recall that we assume $m_u=0$; using a single large constituent 
mass to describe the kinematics should be sufficient for us). 
Fig.~\ref{lifebrs} shows
that this prescription implies a switch--over from Eq.~(\ref{en2}) to
Eq.~(\ref{ea1}) at $\Delta m_{\tilde \chi} \simeq 1.5$ GeV.

\section*{Appendix B: The LSP Relic Density}
\renewcommand{\theequation}{B.\arabic{equation}} 
\setcounter{equation}{0}
Our calculation of the present mass density of LSP's left over from
the Big Bang follows the treatment of Ref.~\cite{griest}. The physical
picture is that the LSP remains in thermal equilibrium until the
Universe has cooled to the temperature $T_F$ where the LSP ``freezes
out''. At lower temperatures essentially no further LSP's are produced, but
occasionally two of them still annihilate into standard model
particles, thereby reducing the relic density. In our case we have to
deal with the additional complication \cite{mngy} that the LSP might
be almost degenerate in mass with the lightest chargino.  In this case
reactions that convert a neutralino into a chargino or vice versa,
such as $\tilde{\chi}_1^0 + f \leftrightarrow \tilde{\chi}_1^\pm +
f'$, remain in thermal equilibrium long after the LSP density itself
has dropped out of equilibrium. The reason is that the rate for such
conversion reactions is proportional to the product of the small LSP
density and a large density of some SM fermion, whereas the rate for
LSP annihilation processes is proportional to the {\em square} of the
small LSP relic density. The chargino density is therefore simply
given by the neutralino density times a Boltzmann factor. One can then
include co--annihilation effects by means of an effective LSP
annihilation cross section.

Following Ref.~\cite{griest}, we first define the effective number of LSP 
degrees of freedom,
\begin{equation} \label{eb1}
g_{\rm eff} = 2 + 4 \left( 1 + \Delta_{\tilde \chi} \right)^{3/2}
\exp(-\Delta_{\tilde \chi} x),
\end{equation}
where we have introduced $\Delta_{\tilde \chi} \equiv m_{\tilde \chi_1^\pm}
/ m_{\tilde \chi_1^0} - 1$ and the inverse rescaled temperature $x \equiv
m_{\tilde \chi_1^0} /T$. Notice that the LSP, being a Majorana fermion, only
has two degrees of freedom, whereas the chargino has four. However, the
contribution of the chargino is suppressed by the Boltzmann factor
$\exp(-\Delta_{\tilde \chi} x)$. The effective annihilation cross section is
then given by
\begin{equation} \label{eb2}
\sigma_{\rm eff} = \frac {4} {g_{\rm eff}^2} \sigma (\tilde{\chi}_1^0
\tilde{\chi}_1^0 \rightarrow {\rm anything}) + \frac{16} {g^2_{\rm eff}}
\left( 1 + \Delta_{\tilde \chi} \right)^{3/2} \exp(-\Delta_{\tilde \chi} x)
\sigma (\tilde{\chi}_1^0 \tilde{\chi}_1^- \rightarrow {\rm anything}).
\end{equation}
Notice the relative factor of 4 between the first and the second term
in Eq.~(\ref{eb2}). One factor of two arises because of the larger number
of chargino degrees of freedom, and another factor of two appears
because here the initial state contains two different particles
\cite{griest}. In principle we should also add terms for
$\tilde{\chi}_1^+ \tilde{\chi}_1^-$ annihilation and, if $m_{\tilde
\chi_1^-} > m_W$, for $\tilde{\chi}_1^- \tilde{\chi}_1^-$ and
$\tilde{\chi}_1^+ \tilde{\chi}_1^+$ annihilation. However, it turns
out that the terms already included in Eq.~(\ref{eb2}) are sufficient
to reduce the relic density to a value that is too small to be of
cosmological significance if the chargino--neutralino mass splitting
is small, see Fig.~16; the exact value of the relic density is then of
little interest. The relic density only reaches significant levels
if the Boltzmann factor is already much smaller than 1. In this case
the terms we have omitted are very small, since they are suppressed by
the square of this factor. 

Numerically, freeze--out occurs ar $x=x_F \simeq 20$. Since $x_F \gg
1$, the LSP's and charginos are quite nonrelativistic when they drop
out of thermal equilibrium. Further, the low freeze--out temperature
means that the co--annihilation contribution to the effective cross
section (\ref{eb2}) starts to become suppressed if the LSP--chargino
mass difference exceeds a few percent. Nevertheless in our case the
second term remains dominant out to quite large mass splittings, because
the $\tilde{\chi}_1^0 \tilde{\chi}_1^0 $ annihilation term is strongly
suppressed: since $\tilde{\chi}_1^0$ is an almost pure gaugino state,
its couplings to $Z$ and Higgs bosons are small; further, sfermion
exchange contributions are suppressed by the large sfermion masses
required in this model, unless $|\delta_{GS}|$ is either much smaller or
much larger than 1. In contrast, the $\tilde{\chi}_1^0 \tilde{\chi}_1^- $
annihilation cross section is quite large here, since the 
$\tilde{\chi}_1^0 \tilde{\chi}_1^- W^+$ coupling is near its maximum in
our case.

We computed the first term in Eq.~(\ref{eb2}) including the full set of
2--body final states treated in Ref.~\cite{dn1}. We used the usual
nonrelativistic expansion of the cross sections in most cases, but
treated the thermal average over Breit--Wigner factors due to
$s-$channel exchange of Higgs and $Z$ bosons more carefully
\cite{griest}, using a numerical method developed in Ref.~\cite{dy1}. 
In the second term of Eq.~(\ref{eb2}) we only included $f \bar{f}'$ and
$W \gamma$ final states, where $f$ and $f'$ are light SM fermions, whose
masses we neglected. Since sfermions as well as the charged Higgs bosons
are very heavy in the model we are studying, we only included contributions
from the exchange of $W$ bosons and (for the $W \gamma$ final state) the
light chargino. As discussed in the main text, present experimental bounds
imply that $m_{\tilde \chi_1^-} + m_{\tilde \chi_1^0}$ is well above
$m_W$. We can therefore use the nonrelativistic expansion when calculating
the co--annihilation cross sections. The result is:
\begin{eqnarray} \label{eb3}
\sigma( \tilde{\chi}_1^0 \tilde{\chi}_1^- \rightarrow f \bar{f}') =
\frac {N_cg^4}{64 \pi} & & \frac {1} { \left( s - m_W^2 \right)^2 }
\biggl\{ s \left( O_{11}^R + \eta O^L_{11} \right)^2 + v^2 O^R_{11}
O^L_{11} \widetilde{m}_0 \widetilde{m}_- \\
& &
+ \frac {v^2}{2} \left[ \left( O_{11}^R \right)^2 +
\left( O_{11}^L \right)^2 \right] \left[ \frac{5}{3}  |\widetilde{m}_0| 
\widetilde{m}_- + \widetilde{m}_-^2 + \widetilde{m}_0^2 \right] \biggr\}. 
\nonumber 
\end{eqnarray}
\begin{eqnarray} \label{eb4}
\sigma( \tilde{\chi}_1^0 \tilde{\chi}_1^- \rightarrow W^- \gamma) =
\frac { \alpha_{\rm em} g^2} {8 s} 
\Biggl\{ & & \left[ \left( O_{11}^R \right)^2 +
\left( O_{11}^L \right)^2 \right] \left[ 1 + \left( \frac {\widetilde{m}_0}
{\widetilde{m}_-} \right)^2 \right] \left( 4 + \gamma_W^2 \right) 
\nonumber \\ & &
- 4 \gamma_W^2 O^L_{11} O^R_{11} \frac {\widetilde{m}_0} {\widetilde{m}_-}
\Biggr\}.
\end{eqnarray}
Here, $g$ is the $SU(2)$ gauge coupling, $N_c, \ O^{R,L}_{11}, \ 
\widetilde{m}_0$ and $\widetilde{m}_-$ have been defined after Eq.~(A.1),
$v$ is the relative velocity between the chargino and the neutralino in
their center--of--mass frame, and $s \equiv \left( |\widetilde{m}_0| +
\widetilde{m}_- \right)^2$. Notice that, following Ref.~\cite{dn1}, we
are working in a convention where the neutralino mixing matrix is real;
in that case we must allow the neutralino masses to have either sign, and
$\eta$ in Eq.~(\ref{eb3}) is the sign of $\widetilde{m}_0$. The chargino
masses can always chosen to be positive even if we take the chargino mixing
matrices to be real. Finally, $\gamma_W = E_W/m_W = (s+m_W^2)/(2 m_W^2)$.
The results of eqs.(\ref{eb3}),(\ref{eb4}) already include summing and
averaging over spin and helicity states; the terms $\propto \gamma_W^2$ in
Eq.~(\ref{eb4}) are due to the production of longitudinal gauge bosons.
Note that these terms cancel in the limit $\widetilde{m}_0 = 
\widetilde{m}_-, \ O_{11}^L = O_{11}^R$; this ensures that the cross section
drops like $1/s$ in the limit of large $SU(2)$ gaugino mass $M_2$, as
required by unitarity. Finally, since the contribution from Eq.~(\ref{eb4})
turns out to be at least a factor of 10 smaller than that from Eq.~(\ref{eb3}),
we have only computed the leading ${\cal O}(v^0)$ term here.

\clearpage
 

\begin{thebibliography}{99}
\frenchspacing
\small

\bibitem{nonu} A. Lleyda and C. Munoz, \PLB B317 82 1993 ;
D. Choudhury, F. Eberlein, A. Konig, J. Louis, and S. Pokorski,
\PLB B342 180 1995 ;
M. Olechowski and S. Pokorski, \PLB B344 201 1995 ;
N. Polonsky and A. Pomarol, \PRD D51 6532 1995 ;
D. Matalliotakis and H.P. Nilles, \NPB B435 115 1995 ;
P. Brax, U. Ellwanger and C.A. Savoy, \PLB B347 269 1995 ;
S. Dimopoulos and G.F. Giudice, \PLB B357 573 1995 ;
T. Kobayashi \etal, \PLB B348 402 1995 ;
R. Altendorfer and T. Kobayashi, \IJMP A11 903 1996 .


\bibitem{ibanez} A. Brignole, L.E. Ibanez, and C. Munoz, \NPB B422 125 1994 ,
Erratum, \ibid, {\bf B436} 747 (1995).

\bibitem{ibanezii} A. Brignole, L.E. Ibanez, C. Munoz, and C. Scheich,
FTUAM-95-26, hep-ph/9508258.

\bibitem{clm} J.A. Casas, A. Lleyda and C. Munoz, FTUAM 96/03, hep-ph/9601357.

\bibitem{gmmech} G.F. Giudice and A. Masiero, \PLB B206 480 1988 .

\bibitem{hhg} J.F. Gunion, H.E. Haber, G.L. Kane, and S. Dawson,
{\it The Higgs Hunters Guide}, Addison Wesley.

\bibitem{mngy} S. Mizuta, D. Ng, and M. Yamaguchi, \PLB B300 96 1993 .

\bibitem{pappierce} A. Papadopoulos and D. Pierce, \NPB B430 278 1994 .

\bibitem{l3recent} The L3 Collaboration, CERN-PPE/96-29 (1996).

\bibitem{cdgi} C.-H. Chen, M. Drees, and J.F. Gunion, \PRL 76 2002 1996 .

\bibitem{bartl} A. Bartl, H. Fraas and W. Majerotto, \ZP C30 441 1986 ;
\NPB B278 1 1986 .

\bibitem{tbarklow}  T. Barklow, private communication.

\bibitem{D0cuts} D0 Collaboration, S. Abachi \etal, \PRL 75 618 1995 .

\bibitem{CDFcuts} CDF Collaboration, J. Hauser, Fermilab-Conf-95/172-E,
published in {\it Proceedings of the 10th Topical Workshop on Proton-Antiproton
Collider Physics}, American Institute of Physics.


\bibitem{baeretal} H. Baer, C.-H. Chen, F. Paige, and X. Tata, FSU-HEP-960216.

\bibitem{baeretalold} H. Baer, C.-H. Chen, F. Paige, and X. Tata,
\PRD D52 2746 1995 .

\bibitem{tkamon} T. Kamon, CDF/PUB/EXOTIC/PUBLIC/3667,
presented at the XXXI Rencontre de Moriond, May 1996.

\bibitem{ATLAS} ATLAS Technical Proposal, CERN/LHCC/94-43, LHCC/P2 (1994).

\bibitem{CMS} CMS Technical Proposal, CERN/LHCC 94-38, LHCC/P1 (1994).

\bibitem{womersley} J. Womersley, private communication.

\bibitem{lopezetal} 
P. Nath and R. Arnowitt, \PRL 70 3696 1993 , and \PLB B299 58 1993 ~[erratum:
{\it ibid.} {\bf B307}, 403 (1993)];
J.L. Lopez, D.V. Nanopoulos, Ka-jia Yuan, \PRD D48 2766 1993 ;
R. Arnowitt and P. Nath, hep-ph/9509260 and hep-ph/9411350.

\bibitem{kanehaber}
H.E. Haber and G.L. Kane, Phys. Rep. {\bf 117}, 75 (1985).
\bibitem{kuhn}
H. K\"uhn and A. Santamaria, Z. Phys. {\bf C48}, 445 (1990).
\bibitem{griest}
K. Griest and D. Seckel, Phys. Rev. {\bf D43}, 3191 (1991).
\bibitem{dn1}
M. Drees and M.M. Nojiri, Phys. Rev. {\bf D47}, 376 (1993).
\bibitem{dy1}
M. Drees and A. Yamada, Phys. Rev. {\bf D53}, 1586 (1996).

\end{thebibliography}
\end{document}